\documentclass[aps,pre,reprint,twocolumn,superscriptaddress,showpacs]{revtex4-2}
\usepackage{amssymb,amsmath}
\usepackage[table]{xcolor}
\usepackage{graphicx}
\usepackage{mathtools}
\usepackage[export]{adjustbox}
\usepackage{overpic}
\usepackage[colorlinks=true,
 linkcolor=black,
 urlcolor=blue,
 citecolor=blue]{hyperref}
\usepackage{nicefrac}
\usepackage{multirow}
\usepackage{lipsum} 
\usepackage[normalem]{ulem}
\usepackage{pbox}
\newcolumntype{C}[1]{>{\centering\let\newline\\\arraybackslash\hspace{0pt}}m{#1}}
\usepackage{verbatim}
\usepackage{enumitem}

\usepackage{framed,color}
\definecolor{shadecolor}{rgb}{0.85,0.80,0.80}

\definecolor{myorange}{RGB}{253, 184, 99}
\definecolor{mypurple}{RGB}{178, 171, 210}

\newcommand{\comments}[1]{}
\usepackage{multirow}

\usepackage{float}

\renewcommand{\vec}[1]{\mathbf{#1}}

\newcommand{\vx}{\vec{x}}
\newcommand{\vp}{\vec{p}}

\renewcommand{\vec}[1]{\mbox{\boldmath$#1$}}

\newcommand{\beq}{\begin{equation}}
\newcommand{\eeq}{\end{equation}}
\newcommand{\bal}{\begin{aligned}}
\newcommand{\eal}{\end{aligned}}

\newcommand{\be}{\begin{equation}}
\newcommand{\ee}{\end{equation}}
\newcommand{\bd}{\begin{displaymath}}
\newcommand{\ed}{\end{displaymath}}
\newcommand{\BE}{\begin{eqnarray}}
\newcommand{\EE}{\end{eqnarray}}

\newcommand{\id}{{\openone}}

\newcommand{\bx}{\ensuremath{\mathbf{x}}}

\newcommand{\boldpsi}{{\mbox{\boldmath $\psi$}}}

\newcommand{\jb}[1]{\textbf{\textcolor{green}{[JB: #1]}}}
\allowdisplaybreaks

\begin{document}
\title{Eigenvalues of random matrices with generalised correlations: \\ a path integral approach}
\author{Joseph W. Baron}
\email{josephbaron@ifisc.uib-csic.es}
\affiliation{Instituto de F{\' i}sica Interdisciplinar y Sistemas Complejos IFISC (CSIC-UIB), 07122 Palma de Mallorca, Spain}
\author{Thomas Jun Jewell}
\affiliation{Department of Physics and Astronomy, School of Natural Sciences,
	The University of Manchester, Manchester M13 9PL, United Kingdom}
\author{Christopher Ryder}
\affiliation{Department of Physics and Astronomy, School of Natural Sciences,
	The University of Manchester, Manchester M13 9PL, United Kingdom}
\author{Tobias Galla}
\email{tobias.galla@ifisc.uib-csic.es}
\affiliation{Instituto de F{\' i}sica Interdisciplinar y Sistemas Complejos IFISC (CSIC-UIB), 07122 Palma de Mallorca, Spain}
\affiliation{Department of Physics and Astronomy, School of Natural Sciences,
	The University of Manchester, Manchester M13 9PL, United Kingdom}
 
\begin{abstract}
Random matrix theory allows one to deduce the eigenvalue spectrum of a large matrix given only statistical information about its elements. Such results provide insight into what factors contribute to the stability of complex dynamical systems. In this letter, we study the eigenvalue spectrum of an ensemble of random matrices with correlations between any pair of elements. To this end, we introduce an analytical method that maps the resolvent of the random matrix onto the response functions of a linear dynamical system. The response functions are then evaluated using a path integral formalism, enabling us to make deductions about the eigenvalue spectrum. Our central result is a simple, closed-form expression for the leading eigenvalue of a large random matrix with generalised correlations. This formula demonstrates that correlations between matrix elements that are not diagonally opposite, which are often neglected, can have a significant impact on stability. 
\end{abstract}

\maketitle

Determining the factors that contribute to the stability of a dynamical system with many interacting components is a fundamental problem. The theory of large random matrices demonstrates that we can ascertain the stability of such a system given only statistical information about its microscopic interactions. As such, random matrix theory (RMT) has found myriad applications outside its original field of conception, that of nuclear and atomic physics \cite{wigner1958distribution, wigner1967random}, and has become a rich and active area in its own right \cite{tao2012topics, mehta2004random, anderson2010introduction,Vivo, bouchaud}. Among the diverse range of fields where RMT enjoys a centrally important role are spin glasses \cite{mezard1987, braymoore, kosterlitz1976spherical, de1983eigenvalues}, complex ecosystems \cite{may, allesinatang1, allesinatang2,  gravel, barabas2017} and neural networks \cite{ aljadeff2015transition, coolen_kuehn_sollich, kuczala2016eigenvalue, rajan2006eigenvalue, louart2018random}. 

As a result of the desire to apply RMT in a greater range of contexts, its remit has been expanded to encompass an evermore complete collection of random matrix ensembles. For example, it has been shown that Wigner's semi-circle law \cite{wigner1958distribution, wigner1967random} can be generalised for asymmetric matrices, which have eigenvalues that are uniformly distributed in an ellipse in the complex plane \cite{girko1986elliptic, sommers}. Allowing for a uniform non-zero mean for each of the matrix elements gives rise to an additional outlier eigenvalue \cite{tao2013outliers, edwardsjones, orourke, feral2007largest}. Recently, the eigenvalue spectra of more elaborate block-structured matrices \cite{grilli2017,baron2020dispersal} and matrices with element-specific variability \cite{aljadeff2015transition, kuczala2016eigenvalue} have also been investigated.

Despite the aforementioned developments, typically only correlations between matrix elements that are diagonally opposite each other (i.e., $a_{ij}$ and $a_{ji}$) are included in RMT calculations (with the notable exception of Ref. \cite{aceitunorogersschomerus}, where cyclic correlations are explored). This is a rather artificial restriction. In fact, correlations between matrix elements sharing only one index (e.g. $a_{ij}$ and $a_{ki}$) have been shown to arise organically in dynamically evolved ecosystems \cite{bunin2016interaction}, and they appear in contexts as wide-ranging as statistical inference \cite{allen2012inference}, game theory \cite{berg1998matrix,galla2012relative} and data security~\cite{kasiviswanathan2010price}.  

The main result of this letter is a surprisingly simple closed-form expression for the leading eigenvalue of an ensemble of large random matrices with generalised correlations. This includes correlations between elements that are not transpose pairs. The formula that we derive demonstrates directly that such correlations can have a significant impact on stability and therefore should not be dismissed. 

To obtain this result, we have developed a new analytical approach. We exploit a correspondence between the resolvent of the random matrix and the response functions of a linear dynamical system \cite{cui2020perturbative}. With this duality in mind, one can write down the Martin-Siggia-Rose-Janssen-de Dominicis (MSRJD) \cite{msr, janssen1976lagrangean, dedominicis,  altlandsimons, hertz2016path} path integral for the dynamical system and use field-theoretic methods to find the response functions. We are thus able to find the resolvent matrix and, consequently, the leading eigenvalue.

Our method has several advantages. First, we are able to demonstrate explicitly that our results do not depend on the precise distribution from which the matrix elements are drawn (a property known as universality \cite{taovu2010, taovukrishnapur2010}). Further, the dynamical approach we use has no need for replicas \cite{dedominicis1978dynamics}. Finally, we are able to include the effects of each of the different correlations in our ensemble one-by-one, greatly reducing the complexity of the calculation. Particularly because of this last property, we believe that our analytical approach would lend itself to simplifying other similarly challenging problems in RMT.

Let us now define more precisely the ensemble of random matrices on which we will focus. Consider a square matrix of size $N\times N$ whose elements $\{a_{ij}\}$ are drawn from a joint probability distribution. We decompose the matrix $\underline{\underline{a}}$ as follows
\begin{align}
a_{ij} = -\delta_{ij} + \frac{\mu}{N} + z_{ij}, \label{adeff}
\end{align}
where $\mu/N$ is the mean of the off-diagonal elements and the $\{z_{ij}\}$ encode fluctuations about this mean such that $\langle z_{ij} \rangle = 0$ (angular brackets indicate an average over realisations of the random matrix). If $\underline{\underline{a}}$ were the Jacobian of a system linearised about a fixed point, the term $-\delta_{ij}$ in Eq.~(\ref{adeff}) would ensure the stability of the system in absence of interactions. Upon including the interactions encoded by the $\{z_{ij}\}$ and $\mu/N$, one is able to deduce what statistical features of the interactions between components tend to make the system unstable.

We consider the most generic set of pairwise correlations for the elements $z_{ij}$ that do not privilege any position in the matrix over any other. These are 
\begin{align}
\mathrm{Var}(z_{ij}) &= \frac{\sigma^2}{N},\nonumber \\
\mathrm{Corr}(z_{ij}, z_{ji}) &= \Gamma , \nonumber \\
\mathrm{Corr}( z_{ij}, z_{ki}) &= \frac{\gamma}{N}, \nonumber \\
\mathrm{Corr}( z_{ij}, z_{ik}) &= \frac{r }{N}, \nonumber \\
\mathrm{Corr}( z_{ji}, z_{ki}) &= \frac{c }{N}, \label{zcorr}
\end{align}
where none of the indices $i$, $j$, or $k$ take equal values, and where we use the shorthand $\mathrm{Corr}(a,b) = [\langle a b \rangle - \langle a\rangle \langle  b\rangle]/\sqrt{\mathrm{Var}(a)\mathrm{Var}(b)}$ for the correlation coefficients. 

The scaling of the mean and the variance of each element with $N$ in Eqs.~(\ref{adeff}) and (\ref{zcorr}) ensures a sensible thermodynamic limit \cite{edwardsjones, mezard1987, sommers}. 

We distinguish between two types of correlations in Eq.~(\ref{zcorr}): type-1 correlations between diagonally opposite elements [these correlations are ${\cal O}(N^0)$ and quantified by $\Gamma$], and type-2 correlations between elements with one index in common (whose magnitudes are governed by $\gamma$, $c$ and $r$). The type-2 correlations must be of the order $N^{-1}$, again in order to ensure a sensible thermodynamic limit \cite{bunin2016interaction, galla2012relative, berg1998matrix}. 

We do not consider explicitly correlations between elements that share no indices; the presence of such correlations can be shown to be produced by a value of $\mu$ that fluctuates between realisations of the random matrix (see Section I.B of the Supplemental Material (SM) \footnote{The Supplemental Material contains further details of the calculation of both the outlier eigenvalue and bulk eigenvalue density, as well as a method for the numerical production of random matrices with generalised correlations.}). 
\begin{figure}[h]
	\centering 
	\includegraphics[scale = 0.48]{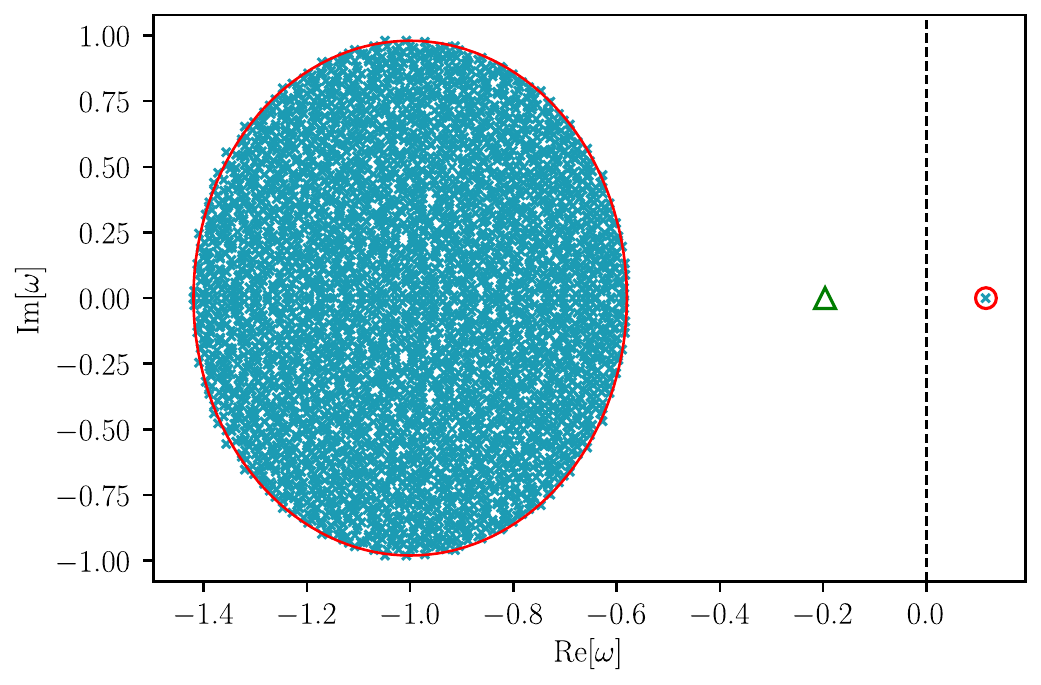}
	\caption{An example eigenvalue spectrum. Crosses represent the results of numerical diagonalisation of computer-generated random matrices from the ensemble with generalised correlations. The solid line is the ellipse $(1+ x)^2/(1+\Gamma)^2 + y^2/(1-\Gamma)^2 = \sigma^2$ where $\omega = x + i y$. The red circle is the prediction for the outlier from Eq.~(\ref{outlierexpression}). The green triangle is the prediction one would obtain ignoring type-2 correlations [$\lambda_{\Delta}=-1+\mu +\Gamma\sigma^2/\mu$]. Parameters are: $N = 4000$, $\sigma = 0.7$, $\Gamma = -0.4$, $\mu = 1$, $\gamma = 0.7$, $r = 1.4$, $c = 1.6$. }\label{fig:examplespectrum}
\end{figure}

We now turn our attention to the eigenvalue spectrum. Consider the disorder-averaged eigenvalue density, which is defined as
\begin{align}
\rho(\omega) = \left\langle \frac{1}{N} \sum_{i = 1}^N \delta(\omega -\lambda_i) \right\rangle, \label{eigenvaluedensity}
\end{align}
where the $\{\lambda_i\}$ are the eigenvalues of any one realisation of the random matrix. The eigenvalue density is normalised such that $\int d^2\omega \,\rho(\omega) =1$, where the integral covers the entire complex plane. 

The non-zero mean of the elements of the matrix $\underline{\underline{a}}$ constitutes a rank-1 perturbation to the matrix $-\underline{\underline{\id}} + \underline{\underline{z}}$ \cite{tao2013outliers, orourke, feral2007largest, edwardsjones}. Previous work thus allows us to anticipate the general form that the eigenvalue density $\rho(\omega)$ will take \cite{edwardsjones, orourke, allesinatang2}. There are two contributions to the eigenvalue spectrum: a bulk region, to which the vast majority of the eigenvalues are confined, and a single outlier that results from the non-zero mean. An example is shown in Fig. \ref{fig:examplespectrum}. More precisely, we write
\begin{align}
\rho(\omega) = \rho_{\mathrm{bulk}}(\omega) + \frac{1}{N} \delta(\omega - \lambda_{\mathrm{outlier}}) .
\end{align}

The primary tool that we use for calculating the outlier eigenvalue and the bulk eigenvalue density is the resolvent matrix, defined in our case by
\begin{align}
\underline{\underline{G}} = \left\langle \left[(1 + \omega)\underline{\underline{\id}} - \underline{\underline{z}}\right]^{-1}\right\rangle.\label{resolventdef}
\end{align}
The bulk eigenvalue density can be calculated from only the trace of the resolvent matrix. We define $G(\omega, \omega^\star) = \frac{1}{N} \sum_i G_{ii}(\omega, \omega^\star)$, noting that the resolvent may not necessarily be an analytic function of $\omega$. The bulk spectrum can then be obtained as \cite{sommers}
\begin{align}
\rho_{\mathrm{bulk}}(\omega) = \frac{1}{2\pi}\mathrm{Re}\left[\frac{\partial G}{\partial \omega^\star}\right] . \label{bulkfromresolvent}
\end{align}
Using established techniques \cite{sommers, haake, anandgalla, hubbard}, we find (see Section III of the SM) that the bulk eigenvalue density is independent of $r$, $c$ and $\gamma$ (i.e. the type-2 correlations), and is given by the familiar elliptical law \cite{sommers, girko1986elliptic}  
\begin{align}
\rho_{\mathrm{bulk}}(\omega) = 
\begin{cases}
[\pi \sigma^2 (1-\Gamma^2)]^{-1} \,\,\,\,\, \mathrm{if} \,\,\,\,\, \frac{(1+ x)^2}{(1+\Gamma)^2} + \frac{y^2}{(1-\Gamma)^2 }< \sigma^2, \\
0 \,\,\,\,\, \mathrm{otherwise},
\end{cases}\label{bulkspectrum}
\end{align}
where $\omega = x+iy$. This is verified in Fig. \ref{fig:examplespectrum}. 

We now turn our focus to finding the outlier eigenvalue $\lambda_{\mathrm{outlier}}$, for which the effect of the type-2 correlations is significant. The outlier eigenvalue satisfies the known relation \cite{orourke, BENAYCHGEORGES2011494} (see also Section II.A of the SM)
\begin{align}
1 - \frac{\mu}{N} \sum_{ij} G_{ij}(\lambda_{\mathrm{outlier}}) =0 . \label{outlierfromresolvent}
\end{align}
Notably, the off-diagonal elements of the resolvent matrix are required to find the outlier. Often, the resolvent matrix is presumed to be diagonal \cite{braymoore, kuczala2016eigenvalue, hertz2016path, baron2020dispersal}, but in the more general case considered here, the off-diagonal elements turn out to be crucial. 

Two observations aid us in evaluating the resolvent matrix, including its off-diagonal elements: (i) The resolvent is an analytic function of $\omega$ outside the bulk region of the eigenvalue spectrum [see Eq.~(\ref{bulkfromresolvent}) and Refs. \cite{sommers, kuczala2016eigenvalue, feinberg1997non}]; (ii) When the resolvent is analytic, we can show that the elements of the resolvent matrix correspond to the response functions of the following linear dynamical system,
\begin{align}
\dot x_i = - x_i + \sum_j z_{ij} x_j + h_i,\label{linearprocess}
\end{align}
where the $\{h_i(t)\}$ are external fields. That is, the Laplace transforms of the response functions $R_{ij}(t- t') = \left\langle \frac{\delta x_i(t)}{\delta h_j(t')} \right\rangle$ (where we exploit time-translation invariance) are equal to the elements of the resolvent matrix defined in Eq.~(\ref{resolventdef}),
\begin{align}
\hat R_{ij}(u) = G_{ij}(u) . \label{responseandresolvent}
\end{align}
This observation is the basis for our calculation. It means that if we can calculate the response functions of the dynamics in Eq.~(\ref{linearprocess}), we can use Eq.~(\ref{outlierfromresolvent}) to find $\lambda_{\mathrm{outlier}}$. 

To find the response functions we begin with the MSRJD generating functional  \cite{msr, janssen1976lagrangean, dedominicis, altlandsimons, hertz2016path} of the system in Eq.~(\ref{linearprocess}), and then carry out the disorder average along the lines of Refs.  \cite{dedominicis1978dynamics, kirkpatrick1987, opper1992phase}. One thus obtains the following path integral expression for the response functions (see Section II.C of the SM)
\begin{align}
R_{ij}(t-t') &= -i\langle x_i(t) \hat x_j(t')\rangle_S, \nonumber \\
\langle \cdots \rangle_S &= \int D[\mathbf{x}, \hat{\mathbf{x}}] \left[\cdots\right] e^{S_0 + S_{\mathrm{int}}}, \label{pathintegral}
\end{align}
where $\int D[\mathbf{x}, \hat{\mathbf{x}}] $ indicates a functional integral over all possible trajectories of the coordinates $\{x_i(t)\}$ and their conjugates $\{\hat x_i(t)\}$ \cite{altlandsimons, hertz2016path}. 

\begin{figure}[H]
	\centering 
	\includegraphics[scale = 0.25]{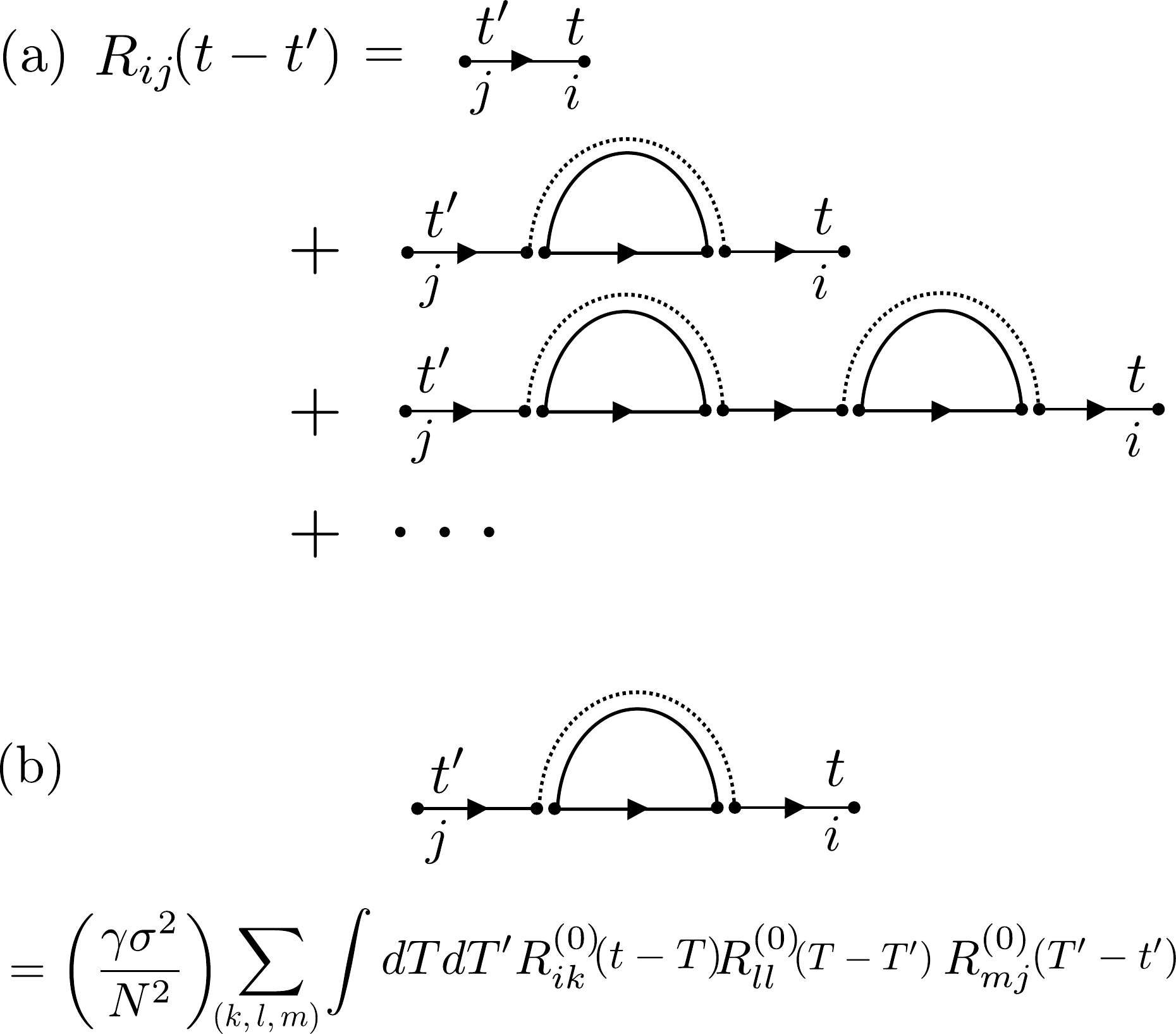}
	\caption{Panel (a) depicts the diagrammatic representation of the series used to evaluate the response function $R_{ij}(t-t')$. Panel (b) indicates the value of the second diagram in panel (a) to leading order in $N^{-1}$. Each node is associated with a site index ($i$, $j$, $k$, etc.). Nodes that are placed closely together have the same time coordinate and each such pair carries a factor of $(\gamma \sigma^2/N^2)^\frac{1}{2}$. Directed edges each carry a factor of the bare response function. Solid undirected arcs connect indices which share the same value and are summed over [$l$ in the example in panel (b)]. Dashed arcs indicate indices that are summed, but cannot take the same value [e.g., $k$ and $m$ in panel (b)].}\label{fig:diagrams}
\end{figure}

We identify two contributions to the action in Eq.~(\ref{pathintegral}):  a `bare' action, which would still be present if we were to set $r = c = \gamma = 0$, and an `interaction' term, which comes about due to the type-2 correlations 
\begin{align}
&S_0 = i\sum_i\int dt \Bigg[\hat x_i(t)\Bigg(\dot x_i(t)+x_i(t)-h_i(t)\Bigg)\Bigg]\nonumber \\ 
&\,\,\,\,\,-\frac{\sigma^2}{2N} \sum_{(i,j)} \int dt dt' \bigg[ \hat x_i(t) \hat x_i(t') x_j(t)x_j(t') \nonumber \\
&\,\,\,\,\,\,\,\,\,\,\,\,\,\,\,+ \Gamma \hat x_i(t) x_i(t') \hat x_j(t')  x_j(t)\bigg], \nonumber \\
&S_{\mathrm{int}} = -\frac{\sigma^2}{2N^2} \sum_{(i,j,k)} \int dt dt' \bigg[ 2 \gamma  \hat x_i(t) x_j(t) \hat x_k(t') x_i(t') \nonumber \\
+& r \hat x_i(t) x_j(t) \hat x_i(t') x_k(t') +  c  \hat x_i(t) x_j(t) \hat x_k(t') x_j(t') \bigg]. \label{action}
\end{align}
The notation $(i,j,k)$ indicates that only combinations where none of indices take equal values contribute to the sum. We note that one benefit of the dynamic formalism presented here is that this action, and therefore the outlier eigenvalue, can be seen to be universal with relative ease \cite{taovu2010, taovukrishnapur2010} (see SM Section II C).


\begin{figure*}
	\centering 
	\includegraphics[scale = 0.45]{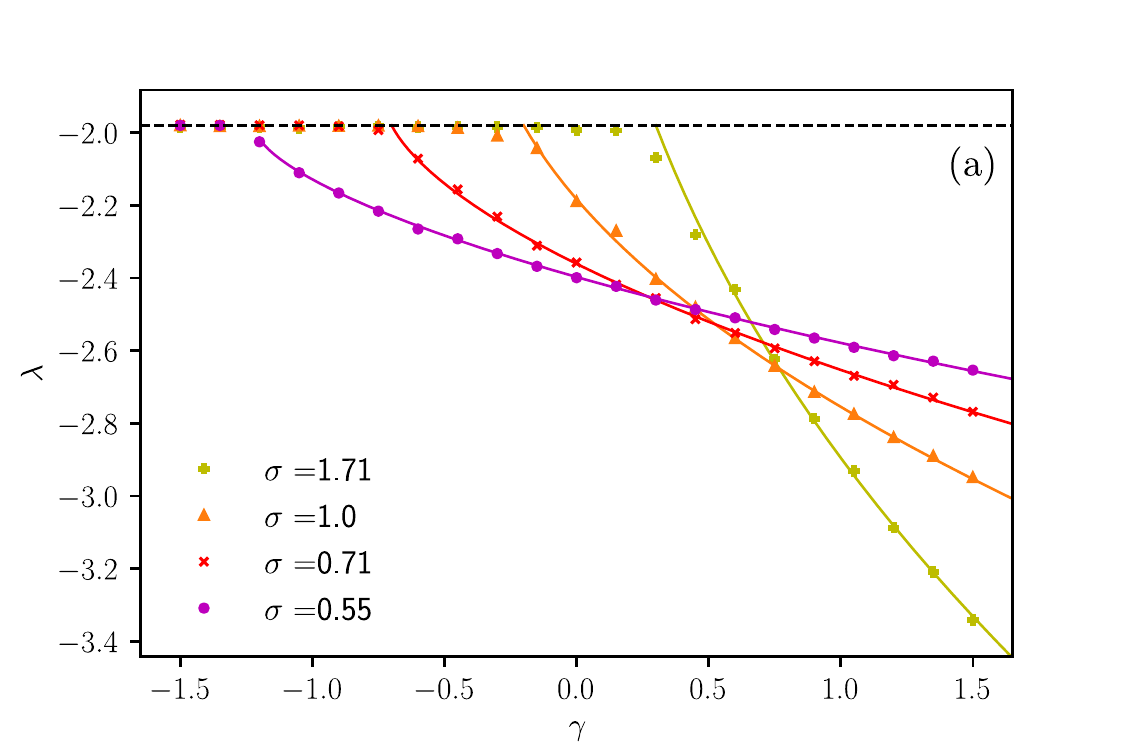}
	\includegraphics[scale = 0.45]{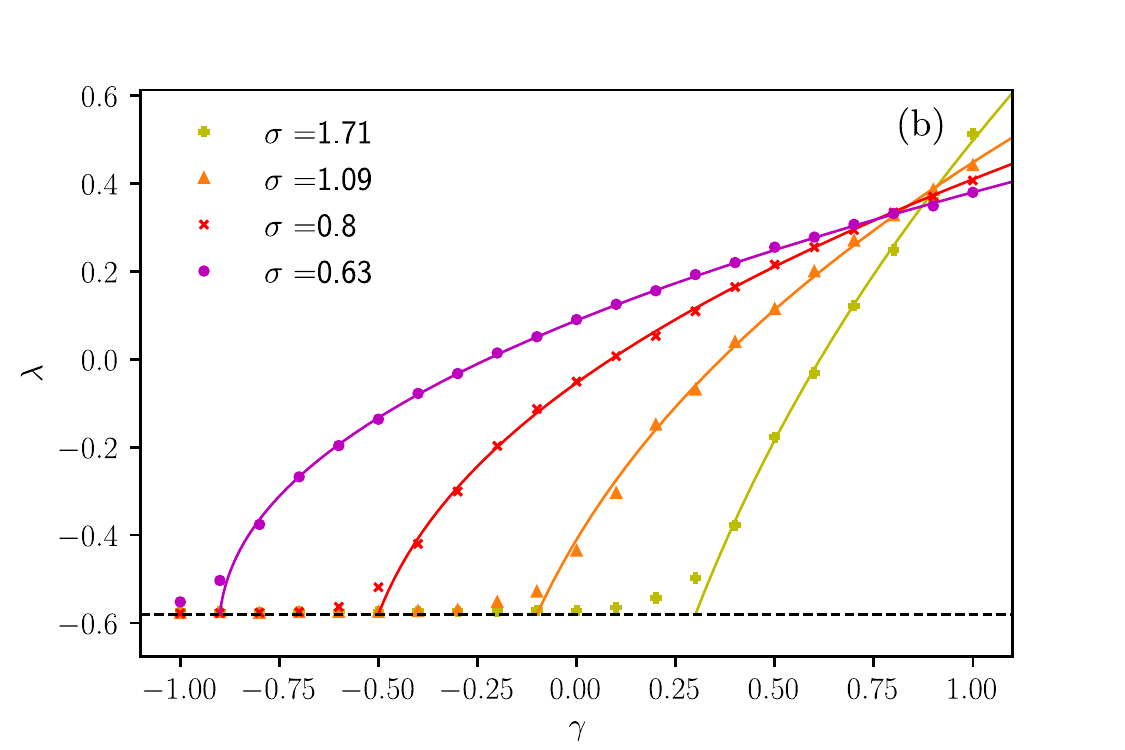}
	\caption{The leftmost eigenvalue [panel (a)] and the rightmost eigenvalue [panel (b)] as a function of the strength of the type-2 correlations $\gamma$ for different combinations of the parameters $\Gamma$ and $\sigma$. Symbols are from numerical diagonalisation of computer-generated random matrices, solid lines are the prediction in Eq.~(\ref{outlierexpression}). The horizontal dashed lines show the edges of the bulk spectrum at $\lambda^\pm_{\mathrm{edge}} = -1 \pm (1+\Gamma)\sigma$. The values of $\sigma$ and $\Gamma$ are such that $(1+\Gamma)\sigma = 0.98$ in panel (a), and $(1+\Gamma)\sigma = 0.42$ in panel (b). In panel (a), matrix entries are from a Bernoulli distribution with $\mu=-1.2$, $r=1.4$, $c=1.6$. In panel (b), matrix entries are constructed from uniformly distributed random numbers such that  $\mu=1.2$, $r=1$, $c=1$. Numerical results are for $N=10000$, averaged over 10 trials. }\label{fig:lambdaversusgamma}
\end{figure*}

We now expand the exponential in Eq.~(\ref{pathintegral}) in powers of $S_{\mathrm{int}}$ and  write the series expansion for the `dressed' response functions $R_{ij}(t-t')$ in terms of the `bare' response functions $R^{(0)}_{ij}(t-t')$ (the response functions of the system with $S_{\mathrm{int}}= 0$). This is accomplished using dynamic mean-field theory, which simultaneously yields an expression for $R^{(0)}_{ij}(t-t')$. Conveniently, our dynamic approach allows us to consider each contribution to the interaction action $S_{\mathrm{int}}$ one-by-one.

The terms of the series for $R_{ij}(t-t')$ can be represented efficiently using diagrams \cite{hertz2016path} (see Fig. \ref{fig:diagrams}). These `rainbow' diagrams have a structure similar to those representing quark-gluon interactions \cite{brezin1994correlation, JANIK1997603}.  

The dynamic formulation of the random matrix problem drastically reduces the complexity of the diagrammatic series that one has to evaluate. This is the main advantage of our approach over established diagrammatic techniques in RMT \cite{JANIK1997603, kuczala2016eigenvalue}. A more detailed discussion can be found in Section II of the SM.


The series expansion of the response function depicted in Fig. \ref{fig:diagrams} can be evaluated exactly in the limit $N\to \infty$, without the need for further approximation. We find
\begin{align}
\frac{1}{N}\sum_{ij}\hat R_{ij}& = \hat R^{(0)} \left[1 - \gamma \sigma^2 (\hat R^{(0)})^2 \right]^{-1}, \label{sellresult}
\end{align}
where $\hat R^{(0)}(u)$ is the average diagonal element of the bare resolvent matrix, given by $\hat R^{(0)}(u) = \frac{(1+u)}{2\Gamma \sigma^2}\left[ 1 + \sqrt{1-4 \Gamma\sigma^2/(1+u)^2}\right]$ for $\vert 1+u \vert >\sigma(1+\Gamma)$ and $u \in \mathbb{R}$. 

Using the correspondence in Eq.~(\ref{responseandresolvent}) we substitute the result of Eq.~(\ref{sellresult}) into Eq.~(\ref{outlierfromresolvent}) and solve for $\lambda_{\mathrm{outlier}}$. We finally obtain our central result: a formula for the outlier eigenvalue 
\begin{align}
\lambda_{\mathrm{outlier}}=
 - 1 + \mu  + \frac{\mu}{2}  \left( 1 + \frac{\Gamma}{\gamma}\right)\left(\sqrt{1 + \frac{4\gamma \sigma^2}{\mu^2} }-1\right). \label{outlierexpression}
\end{align}

\vspace{10pt}

This expression is valid when $\vert\mu \vert> \sigma(1-\gamma)$; there is no outlier eigenvalue when $ \vert\mu \vert\leq \sigma(1-\gamma)$. We observe that $\lambda_{\mathrm{outlier}} \to -1\pm(1+\Gamma) \sigma$ when $\mu \to \pm\sigma(1-\gamma)$. That is, the points at which the expression in Eq.~(\ref{outlierexpression}) becomes invalid correspond to the outlier being absorbed into the bulk region. In the limit $\gamma \to 0$, we recover from Eq.~(\ref{outlierexpression}) the known result $\lambda_{\mathrm{outlier}}= -1 + \mu +\Gamma \sigma^2/\mu$ \cite{edwardsjones, orourke, baron2020dispersal}, which is valid when $\vert \mu \vert> \sigma$.

We note that the in-row and in-column correlations, quantified by $r$ and $c$ respectively, do not appear in Eq.~(\ref{outlierexpression}). As such, these correlations do not alter the eigenvalue spectrum in the thermodynamic limit. This is in contrast to the correlations quantified by $\gamma$, which affect the outlier eigenvalue significantly. The reason for the absence of $r$ and $c$ in Eq.~(\ref{outlierexpression}) is discussed in SM Section II G.

We test the prediction of Eq.~(\ref{outlierexpression}) in Fig.~\ref{fig:lambdaversusgamma} using computer-generated random matrices. The method used to produce random matrices with the correlations in Eqs.~(\ref{zcorr}) is described in Section IV of the SM. 

Fig.~\ref{fig:lambdaversusgamma} not only serves to verify the theoretical predictions in Eqs.~(\ref{outlierexpression}) and (\ref{bulkspectrum}) and their universality, but it also demonstrates the extent to which the type-2 correlations between $a_{ij}$ and $a_{ki}$ (quantified by $\gamma$) can affect the position of the outlier. In particular, the presence of these correlations can give rise to an outlier eigenvalue when there would not ordinarily be one. As is shown in Fig.~\ref{fig:lambdaversusgamma}(b) in particular, this outlier eigenvalue can become positive, resulting in instability, purely as a result of varying $\gamma$ (see also Fig.~\ref{fig:examplespectrum}).

To conclude, correlations between pairs of non-transpose elements (e.g. $a_{ij}$ and $a_{ki}$) are routinely neglected in many problems in random matrix theory. In this work, we developed a dynamic path-integral approach to take these more subtle correlations into account. The diagrammatic series we derived using this approach is far simpler than what one would obtain with established methods. We thus arrived at an explicit formula for the leading eigenvalue of an ensemble of random matrices with correlations between any pair of matrix elements. Hence, we demonstrated directly that such correlations can impact stability and therefore should not be ignored.

It is known that correlations between transpose pairs of interaction coefficients can significantly affect stability in complex ecosystems \cite{ allesinatang2, galla2018dynamically, barbier2018generic, opper1992phase}. However, the effect of the correlations that we study has not been explored in this context (to our knowledge). Interestingly, correlations between non-transpose interaction coefficients can arise quite organically in models of complex ecosystems \cite{bunin2016interaction}. This indicates one immediate opportunity for the application of our results. Additional avenues for future inquiry present themselves in the context of neural networks, where increasingly more general interactions between neurons are being studied \cite{aljadeff2015transition, kuczala2016eigenvalue}. 

We anticipate that the dynamic field-theoretic approach that we developed would also lend itself to solving other challenging problems in random matrix theory. For instance, our method complements replica \cite{semerjiancugliandolo, rodgers1988density, kuhn2008spectra} or cavity techniques \cite{rogers2008cavity, susca2021cavity} used in the calculation of the spectra of sparse random matrices, and would perhaps allow for the generalisation of present results. The calculation of other more complicated ensembles of dense random matrix (such as those with block structure \cite{allesina2015, gravel, grilli2017, baron2020dispersal}) could also be simplified and extended with our approach.

\acknowledgements
We acknowledge partial financial support from the Agencia Estatal de Investigaci\'on (AEI, MCI, Spain) and Fondo Europeo de Desarrollo Regional (FEDER, UE), under Project PACSS (RTI2018-093732-B-C21) and the Maria de Maeztu Program for units of Excellence in R\&D (MDM-2017-0711).

\end{document}


\title{Eigenvalues of random matrices with generalised correlations: \\
	a path integral approach \\ \medskip \medskip{\Large--- Supplemental Material ---}}
\author{Joseph W. Baron}
\email{josephbaron@ifisc.uib-csic.es}
\affiliation{Instituto de F{\' i}sica Interdisciplinar y Sistemas Complejos IFISC (CSIC-UIB), 07122 Palma de Mallorca, Spain}
\author{Thomas Jun Jewell}
\affiliation{Department of Physics and Astronomy, School of Natural Sciences,
	The University of Manchester, Manchester M13 9PL, United Kingdom}
\author{Christopher Ryder}
\affiliation{Department of Physics and Astronomy, School of Natural Sciences,
	The University of Manchester, Manchester M13 9PL, United Kingdom}
\author{Tobias Galla}
\email{tobias.galla@ifisc.uib-csic.es}
\affiliation{Instituto de F{\' i}sica Interdisciplinar y Sistemas Complejos IFISC (CSIC-UIB), 07122 Palma de Mallorca, Spain}
 \affiliation{Department of Physics and Astronomy, School of Natural Sciences,
	The University of Manchester, Manchester M13 9PL, United Kingdom}

\maketitle
\vspace{-2em}
\tableofcontents

\section{Further properties of the random matrix ensemble}\label{section:ensemble}

\subsection{General constraints}\label{section:generalconstraints}
By first considering the quantity $\left\langle \left(z_{ij} \pm z_{ji}\right)^2\right\rangle \geq 0$ and consulting the correlations defined in Eqs.~(2) of the main text, one finds that
\begin{align}
-1 \leq \Gamma \leq 1 .
\end{align}
Further, given that $\left\langle \left (\sum_{j} z_{ij} \pm \sum_k z_{ki}\right)^2\right\rangle \geq 0$, we see that
\begin{align}
\frac{c + r}{2} + 1 \pm (\gamma + \Gamma) \geq 0.
\end{align}
Further constraints on the correlations arise due to the specific choice of numerical method used to generate the random matrices. This is discussed in Section \ref{section:procedure}.

\subsection{Correlations between elements that share no index}\label{section:ftype}
In this section, we examine the possibility of correlations between matrix elements that share no index. That is, we suppose that $\langle z_{ij} z_{kl}\rangle = f \sigma^2/N^2$, where none of $i$, $j$, $k$ or $l$ are equal.

We introduce the following quantity,\begin{align}
\mu^\star = \frac{1}{N}\sum_{ij} a_{ij} \equiv \mu + \xi.
\end{align}
This is the (re-scaled) average matrix element in a single realisation of the random matrix. We have defined the deviation $\xi$ from the ensemble re-scaled mean $\mu$ for a single random matrix realisation. Consider now the variance of $\xi$,
\begin{align}
\langle\xi^2\rangle &= \frac{1}{N^2}\sum_{ijkl} \langle(a_{ij}- \mu/N) (a_{kl}- \mu/N)\rangle \nonumber \\
&\approx \frac{1}{N^2} \left[  N^2\frac{\sigma^2}{N} + N^2\frac{\Gamma\sigma^2}{N} + N^3 \frac{2\gamma \sigma^2}{N^2} + N^3 \frac{r \sigma^2}{N^2} + N^3 \frac{c \sigma^2}{N^2} + N^4\frac{f \sigma^2}{N^2} +\right] \nonumber \\
&= f\sigma^2 + \frac{\sigma^2}{N} \left[ 1 + \Gamma + 2\gamma + r + c\right] . \label{errorvariance}
\end{align}
When $f\neq 0$, we see that the average matrix element varies between realisations of the random matrix, even in the limit $N\to \infty$. These correlations therefore give rise to an effective mean $\mu^\star/N=N^{-2}\sum_{ij}a_{ij}$ that varies between realisations of the matrix. 

However, one notes that we could never measure $f$ from any one instance of the random matrix even in the limit $N\to \infty$.
 Suppose we first measure the mean $\mu^\star/N$ for this matrix and then consider the quantities
\begin{align}
(\sigma^\star)^2 &= \frac{1}{N}\sum_{(i,j)} (a_{ij}- \mu^\star/N)^2, \nonumber \\
f^\star &= \frac{1}{N^2 (\sigma^\star)^2}\sum_{(i,j,k,l)} (a_{ij}- \mu^\star/N) (a_{kl}- \mu^\star/N),
\end{align}
where none of the indices in brackets take equal values in the sums.
We would always find $f^\star = 0$. That is, in the thermodynamic limit, any single matrix drawn from a distribution with statistics $f\neq 0$, $\mu$, $\sigma^2$, $\Gamma$, $\gamma$, $c$ and $r$ is indistinguishable from one drawn from a distribution with statistics $f^\star = 0$, $\mu^\star$, $(\sigma^\star)^2$, $\Gamma^\star$, $\gamma^\star$, $c^\star$ and $r^\star$. There is therefore no need to consider the case $f\neq 0$ if what we desire is to predict the eigenvalue spectrum of any one realisation of the random matrix. 

\section{Calculation of the outlier eigenvalue}\label{section:outlier}
\subsection{Finding the outlier using the resolvent matrix}
The eigenvalues of the random matrix $\underline{\underline{z}} - \underline{\underline{\id}}$ are confined to an elliptic region in the complex plane. This is shown explicitly in Section \ref{section:bulk}. It has been shown previously \cite{tao2013outliers, edwardsjones, orourke, allesinatang1, allesinatang2, baron2020dispersal, BENAYCHGEORGES2011494}, and is again demonstrated in Section \ref{section:independence}, that low-rank perturbations to large random matrices leave the bulk of the eigenvalue spectrum undisturbed, but give rise to the possibility of a small number of outlier eigenvalues. Although the bulk region of the eigenvalue spectrum is unaffected by type-2 correlations, we find that the outlier eigenvalue is highly dependent on these correlations.

We introduce a perturbation to $\underline{\underline{z}} - \underline{\underline{\id}}$ in the form of a uniform matrix with entries $\mu/N$ to produce the matrix $\underline{\underline{a}}$ [see Eq.~(1) in the main text]. The outlier eigenvalue that is possessed by $\underline{\underline{a}}$ but not by $\underline{\underline{z}} - \underline{\underline{\id}}$ satisfies
\begin{align}
\det\left( \underline{\underline{\id}}(1+\lambda) - \underline{\underline{z}} - \frac{1}{N} \underline{\underline{\mu}} \right) =0.
\end{align}
We now define the resolvent matrix
\begin{align}
\underline{\underline{G}}(\omega\vert \{z_{ij}\}) = \left[(1+ \omega)\underline{\underline{\id}} - \underline{\underline{z}}\right]^{-1}. \label{resolventdef}
\end{align}
If we presume that the outlier $\lambda$ of $\underline{\underline{a}}$ resides outside the bulk region, we recover the known result (using Sylvester's determinant identity) \cite{orourke, BENAYCHGEORGES2011494}
\begin{align}
\det\left( \underline{\underline{\mathbb{1}}} - \frac{1}{N} \underline{\underline{\mu}}\, \underline{\underline{G}}(\lambda\vert \{z_{ij}\})\right) =  1 - \frac{\mu}{N} \sum_{ij} G_{ij}(\lambda\vert \{z_{ij}\}) = 0, \label{outliercalcgeneral}
\end{align}
In order to find the outlier eigenvalue, one is tasked with finding the resolvent matrix and solving Eq.~(\ref{outliercalcgeneral}) for $\lambda$. 

One object that appears in Eq.~(\ref{outliercalcgeneral}) is the sum over all elements of the resolvent matrix. In many applications, the resolvent matrix turns out to be diagonal \cite{braymoore, baron2020dispersal} and one often uses this fact in order to simplify its calculation. In our case however, because of the presence of type-2 correlations [the correlations parametrised by $r, c$ and $\gamma$ -- see Eq.~(2) in the main text], the off-diagonal elements of the resolvent matrix turn out to be non-negligible.  

Our strategy for obtaining the resolvent matrix defined in Eq.~(\ref{resolventdef}) is as follows: We first demonstrate the equivalence of the resolvent matrix elements to the response functions of a linear dynamical system. Using established path-integral methods \cite{hertz2016path}, we then determine the response functions of this dynamical system. We show that only the correlations between $z_{ij}$ and $z_{ki}$ (parametrised by $\gamma$) provide non-vanishing contributions and that in-row and in-column correlations (quantified by $r$ and $c$ respectively) make no difference to the value of the outlier.  

\subsection{Equivalence between the resolvent matrix and the response functions for a linear dynamical system}
\noindent Consider the following linear dynamical system for the quantities $x_i(t)$,
\begin{align}
\dot x_i = - x_i + \sum_{j \neq i} z_{ij} x_j + h_i(t), \label{lineareqs}
\end{align}
where $\{h_i(t)\}$ are arbitrary time-varying external fields. Defining response functions $R_{ij}(t,t'\vert \{z_{ij}\}) = \frac{\delta x_{i}(t)}{\delta h_j(t')}$ (for a particular instance of the random matrix), we obtain 
\begin{align}
\frac{\partial}{\partial t} R_{ij}(t,t' \vert \{z_{ij}\}) = -R_{ii}(t,t'\vert \{z_{ij}\}) + \sum_{k \neq i} z_{ik} R_{kj}(t,t'\vert \{z_{ij}\}) + \delta_{ij}\delta(t-t').
\end{align}
Using time translational invariance in the stationary state, taking the Laplace transform and rearranging, we find
\begin{align}
\hat R_{ij}(u\vert \{z_{ij}\}) = \left[(1+u) \underline{\underline{\id}} - \underline{\underline{z}} \right]^{-1}_{ij} = G_{ij}(u\vert \{z_{ij}\}). \label{equivresponseresolvent}
\end{align}
That is, the Laplace transforms of the response functions are the same as the resolvent matrix entries [c.f. Eq.~(\ref{resolventdef})]. 

The above reasoning assumes that the Laplace transform of the response function exists and is an analytic function of $u$. In the region of the complex plane occupied by the bulk of the eigenvalue spectrum of $\underline{\underline{a}}$, the resolvent is no longer analytic \cite{sommers} and the equality in Eq.~(\ref{equivresponseresolvent}) between the Laplace transform of the response functions and the resolvent matrix  no longer holds. However, since we are interested in outlier eigenvalues that lie outside the bulk region of the eigenvalue spectrum, we may still exploit this equivalence.

Following Refs. \cite{sommers, haake, edwardsjones}, we examine the disorder-averaged resolvent\\ $G_{kl}(u) = \int \left(\prod_{ij}dz_{ij}\right) P(\{z_{ij}\}) G_{kl}(u\vert \{z_{ij}\})$. We calculate the corresponding disorder-averaged response functions in the following section.

\subsection{MSRJD path integral and universality}\label{section:msrjd}
To derive the disorder-averaged response functions of our system, we follow Refs.~\cite{altlandsimons, hertz2016path} and construct the MSRJD generating functional \cite{msr, janssen1976lagrangean, dedominicis} for the dynamical system in Eq.~(\ref{lineareqs}) as follows
\BE
Z[\boldpsi,  \mathbf{h} \vert \{ z_{ij}\}] &=&\int D[\bx,\widehat\bx] \exp\Bigg(i\sum_i\int dt \Bigg[\widehat x_i(t)\Bigg(\dot x_i(t)+x(t)-\sum_{j\neq i}z_{ij} x_j(t)-h_i(t)\Bigg)\Bigg]\Bigg)\nonumber\\
&&\times\exp\left(i\sum_i\int dt~ x_i(t)\psi_i(t)\right).\label{eq:gf}
\EE
The response functions can be obtained from the generating functional as derivatives with respect to the source terms and external fields,
\begin{align}
R_{ij}\left(t,t' \vert \{ z_{ij}\}\right) = -i\frac{\delta^2 Z}{\delta \psi_i(t) \delta h_j(t')} . \label{responsefromgenfunct}
\end{align}
Using the statistics in Eq.~(2) of the main text, we average over the disorder in Eq.~(\ref{eq:gf}) to obtain (where we use an overbar instead of angular brackets to denote an average over instances of the random matrix for notational convenience)
\begin{align}
&\overline{\exp\left[ -i\sum_{i \neq j}\int dt z_{ij}\hat x_i(t) x_j(t)\right]} = 1 - \frac{1}{2} \int dt dt' \sum_{i \neq j}\sum_{k \neq l} \overline{z_{ij}z_{kl}}\hat x_i(t) x_j(t)\hat x_k(t') x_l(t') + \cdots\nonumber \\
&= \prod_{i\neq j}\Bigg\{ 1-\frac{1}{2} \int dt dt' \bigg[\frac{\sigma^2}{N} \hat x_i(t) \hat x_i(t') x_j(t)x_j(t') + \frac{\Gamma \sigma^2}{N}  \hat x_i(t) x_i(t') \hat x_j(t')  x_j(t) \nonumber \\
&\,\,\,\,\,  + \frac{ 2 \gamma \sigma^2}{N^2} \sum_{k\neq i, k\neq j} \hat x_i(t) x_j(t) \hat x_k(t') x_i(t') + \frac{ r \sigma^2}{N^2} \sum_{k\neq i, k\neq j} \hat x_i(t) x_j(t) \hat x_i(t') x_k(t')\nonumber \\
&\,\,\,\,\, + \frac{ c \sigma^2}{N^2} \sum_{k\neq i, k\neq j} \hat x_i(t) x_j(t) \hat x_k(t') x_j(t')\bigg] + \cdots\Bigg\}, \label{disorderaverage}
\end{align}
Assuming that the higher order moments of $z_{ij}$ decay more quickly with $N$ than $N^{-1}$, we can truncate the series in the curly brackets and re-exponentiate to obtain the following disorder-averaged generating functional
\begin{align}
\overline{Z[\boldpsi,  \mathbf{h}]} &=\int D[\bx,\widehat\bx] \exp\left(i\sum_i\int dt~ x_i(t)\psi_i(t)\right)\exp\Bigg(i\sum_i\int dt \Bigg[\widehat x_i(t)\Bigg(\dot x_i(t)+x_i(t)-h_i(t)\Bigg)\Bigg]\Bigg)\nonumber\\
&\times\exp\Bigg(-\frac{1}{2} \int dt dt' \bigg[\frac{\sigma^2}{N}\sum_{(i,j)} \hat x_i(t) \hat x_i(t') x_j(t)x_j(t') + \frac{\Gamma \sigma^2}{N} \sum_{(i,j)} \hat x_i(t) x_i(t') \hat x_j(t')  x_j(t) \nonumber \\
&\,\,\,\,\,  + \frac{ \gamma \sigma^2}{N^2} \sum_{(i,j,k)} \hat x_i(t) x_j(t) \hat x_k(t') x_i(t') + \frac{ r \sigma^2}{N^2} \sum_{(i,j,k)} \hat x_i(t) x_j(t) \hat x_i(t') x_k(t')\nonumber \\
&\,\,\,\,\, + \frac{ c \sigma^2}{N^2} \sum_{(i,j,k)} \hat x_i(t) x_j(t) \hat x_k(t') x_j(t')\bigg] \Bigg), \label{genfunctlinearprocess}
\end{align}
where none of the indices in brackets take equal values in the sums. Taking functional derivatives of the disorder-averaged generating functional in Eq.~(\ref{genfunctlinearprocess}) in a similar fashion to Eq.~(\ref{responsefromgenfunct}) yields the disorder-averaged response functions.

We highlight the fact that at no point did we have to assume that the random matrix elements $z_{ij}$ were drawn from any particular distribution. Any deductions we now make about the eigenvalue spectra will therefore apply to any ensemble of random matrices with the statistics in Eq.~(2) of the main text, with the caveat that higher order moments must decay sufficiently quickly with $N$ that the re-exponentiation in Eq.~(\ref{genfunctlinearprocess}) applies. This is known as the universality property \cite{taovu2010, taovukrishnapur2010}.

\subsection{The effective bare action and the series expansion for the dressed response functions}
We now proceed to calculate the disorder-averaged response functions $R_{ij}(t,t')$. From Eq.~(\ref{genfunctlinearprocess}), one reads off the action
\begin{align}
S &= i\sum_i\int dt \Bigg[\widehat x_i(t)\Bigg(\dot x_i(t)+x_i(t)-h_i(t)\Bigg)\Bigg]\nonumber \\ 
&-\frac{1}{2} \int dt dt' \bigg[\frac{\sigma^2}{N}\sum_{(i,j)} \hat x_i(t) \hat x_i(t') x_j(t)x_j(t') + \frac{\Gamma \sigma^2}{N} \sum_{(i,j)} \hat x_i(t) x_i(t') \hat x_j(t')  x_j(t) \nonumber \\
&\,\,\,\,\,  + \frac{2 \gamma \sigma^2}{N^2} \sum_{(i,j,k)} \hat x_i(t) x_j(t) \hat x_k(t') x_i(t') + \frac{ r \sigma^2}{N^2} \sum_{(i,j,k)} \hat x_i(t) x_j(t) \hat x_i(t') x_k(t')\nonumber \\
&\,\,\,\,\, + \frac{ c \sigma^2}{N^2} \sum_{(i,j,k)} \hat x_i(t) x_j(t) \hat x_k(t') x_j(t')  \bigg] . \label{action}
\end{align}
We identify two contributions: (1) a `bare' term which would still be present if we were to set $\gamma = r = c=0$, and (2) an `interaction' term, which results from the correlations between non-transpose pairs of elements. The bare and interaction terms are (respectively)
\begin{align}
S_0 &= i\sum_i\int dt \Bigg[\widehat x_i(t)\Bigg(\dot x_i(t)+x_i(t)-h_i(t)\Bigg)\Bigg]\nonumber \\ 
&-\frac{1}{2} \int dt dt' \bigg[\frac{\sigma^2}{N}\sum_{(i,j)} \hat x_i(t) \hat x_i(t') x_j(t)x_j(t') + \frac{\Gamma \sigma^2}{N} \sum_{(i,j)} \hat x_i(t) x_i(t') \hat x_j(t')  x_j(t) \bigg], \nonumber \\
S_{\mathrm{int}} &= - \int dt dt' \bigg[ \frac{ \gamma \sigma^2}{N^2} \sum_{(i,j,k)} \hat x_i(t) x_j(t) \hat x_k(t') x_i(t') \bigg]\nonumber \\ 
&\,\,\,\,\,  - \frac{1}{2}\int dt dt' \frac{ r \sigma^2}{N^2} \sum_{(i,j,k)} \hat x_i(t) x_j(t) \hat x_i(t') x_k(t') + \frac{ c \sigma^2}{N^2} \sum_{(i,j,k)} \hat x_i(t) x_j(t) \hat x_k(t') x_j(t')  \bigg] . \label{bareandinteraction}
\end{align}
One can show (see Section \ref{appendix:quadraticaction}) that the bare action can be replaced by a statistically equivalent quadratic action in the large-$N$ limit
\begin{align}
S_0^{\mathrm{eff}} =& i\sum_i\int dt \Bigg[\widehat x_i(t)\Bigg(\dot x_i(t)+x_i(t) - \Gamma \sigma^2 \int dt' R_0 (t,t') x_i(t') - h_i(t) \Bigg)\Bigg] \nonumber \\
&-\sum_i \frac{\sigma^2}{2} \int dt dt' C(t,t') \hat x_i(t) \hat x_i(t'). \label{quadraticaction}
\end{align}
Such an action corresponds to the following stochastic process
\BE \label{effectiveprocess}
\dot x_i &= -x_i + \Gamma \sigma^2 \int_0^t dt' R^0(t,t') x_i(t') + \eta_i(t) + h_i(t),
\EE
where $\eta_i(t)$ is a Gaussian noise term, and we have the self-consistency relations
\begin{align} 
R^0(t,t') &\equiv \left \langle \frac{\delta x_i(t)}{\delta h_i(t')} \right\rangle_0, \nonumber \\
C(t,t)&\equiv \langle x_i(t) x_i(t') \rangle_0 = \langle \eta_i(t) \eta_i(t') \rangle_0 .\label{sc}
\end{align}
Here, $\langle  \cdots\rangle_0$ denotes an average over realisations of the noise variables $\{\eta_i(t)\}$, and $R^0(t,t')$ is the bare response function. One notes that these dynamical equations are decoupled for different values of the index $i$ and that each value of $i$ is statistically equivalent. Importantly, this decoupling indicates a diagonal bare resolvent. 

The bare auto-response functions can be found via the functional differentiation of Eqs.~(\ref{effectiveprocess}), followed by an average over realisations of the noise. After taking the Laplace transform in the stationary state, one finds
\begin{align}
\Gamma \sigma^2 (\hat R^0)^2 - (1 + u) \hat R^0 + 1 = 0, \label{bareresponseeq}
\end{align}
which in combination with the constraint $\vert \hat R^0(u) \vert^2< \frac{1}{\sigma^2}$ (see Section \ref{appendix:constraints}) yields the following piecewise expression
\begin{align}
\hat R^0(u) = \begin{cases}
\frac{1}{2\Gamma \sigma^2}\left[ 1 + u + \sqrt{(1+u)^2-4 \Gamma\sigma^2}\right]  \,\,\,\,\,\,\, \mathrm{if} \,\,\,\,\,\,\, 1+u > \sigma(1+\Gamma),\\
\frac{1}{2\Gamma \sigma^2}\left[ 1 + u - \sqrt{(1+u)^2-4 \Gamma\sigma^2}\right]  \,\,\,\,\,\,\, \mathrm{if} \,\,\,\,\,\,\, 1+u <- \sigma(1+\Gamma).
\end{cases}\label{response0}
\end{align}
This can be written in the form given in the main paper,
\be
\hat R^{(0)}(u) = \frac{(1+u)}{2\Gamma \sigma^2}\left[ 1 + \sqrt{1-4 \Gamma\sigma^2/(1+u)^2}\right],
\ee
valid for real $u$ and $|1+u|>\sigma(1+\Gamma)$. 

We note that this agrees with expressions obtained in, for example, Refs. \cite{braymoore, sommers} by other means. Having found the `bare' response functions, we now wish to find the disorder-averaged `dressed' response functions in the case where the correlations quantified by $\gamma$, $r$ and $c$ are non-zero. To do this, we largely follow the procedure in Ref. \cite{hertz2016path}. We first notice that the dressed response functions can be written as an average over paths [see Eq.~(\ref{responsefromgenfunct})] 
\begin{align}
R_{ij}(t,t') = -i\frac{\delta^2 \overline Z}{\delta \psi_i(t) \delta h_j(t')} = -i  \langle  x_i(t) \hat x_j(t')  \rangle_S \label{responsefunctionasaverage}
\end{align}
where we have introduced the notation
\begin{align}
\langle  \cdots \rangle_S = \int D[\bx,\widehat\bx] [\cdots] e^S .
\end{align}
Because we can write the response functions as an average in this way, we can calculate them using a perturbative expansion, with the interaction terms playing the role of the `perturbation'. Such an expansion is arrived at by noticing
\begin{align}
\langle x_i(t) \hat x_j(t')  \rangle_S = \langle x_i(t) \hat x_j(t')  e^{S_{\mathrm{int}}} \rangle_0 = \sum_r \frac{1}{r!} \langle x_i(t) \hat x_j(t')  (S_{\mathrm{int}})^r\rangle_0 , \label{interactionactionexpansion}
\end{align} 
where $\langle  \cdots \rangle_0$ indicates an average with respect to the effective bare action only. Since the effective bare action is quadratic in $x_i(t)$ and $\hat x_j(t)$, Wick's theorem holds for the averages $\langle \cdots \rangle_0$. We can therefore obtain a series that we can evaluate entirely in terms of quantities that are calculable from the non-interacting theory [i.e. from the bare response functions in Eq.~(\ref{response0})].

\subsection{Diagrammatic formalism}\label{section:diagrams} 
We now evaluate the series for the dressed response functions in Eq.~(\ref{interactionactionexpansion}). Diagrams are extremely useful here to keep track of all of the possible Wick pairings. In this section we introduce the diagrammatic convention, following Ref. \cite{hertz2016path}. We note that similar rainbow diagrams were arrived at in a different way in Ref. \cite{JANIK1997603}. We then use these conventions throughout the rest of the calculation.

We first focus on a specific example. Consider the following second-order term in the expansion in Eq.~(\ref{interactionactionexpansion})
\begin{align}
A_{ij}(t,t')=& \frac{1}{ 2!} \left( \frac{\gamma \sigma^2}{N^2} \right)^2 \int dT dT' dT'' dT'''\sum_{(k,l,m)} \sum_{(k',l',m')} \Bigg[ \nonumber \\
&\left\langle x_i(t) \hat x_j(t') \hat x_l(T) x_k(T) \hat x_m(T') x_l(T') \hat x_{l'}(T'') x_{k'}(T'') \hat x_{m'}(T''') x_{l'}(T''')  \right\rangle_0.\Bigg] \label{wickexample}
\end{align}
The average in this expression is found by considering every possible combination of pairs of the Gaussian variables $\{x_a(s)\}$ and $\{\hat x_b(s')\}$. First, we note that pairings such as $\langle \hat x_k(T) \hat x_l(T') \rangle_0$ evaluate to zero, regardless of the values of $k, l$, $T$ and $T'$. This can be seen by considering  
\begin{align}
\langle \hat x_k(T) \hat x_l(T') \rangle_0 = \frac{\partial^2 \overline Z_0[\boldpsi={\mathbf 0},  \mathbf{h}]}{ \partial h_k(t) \partial h_l(T')}  = 0,
\end{align}
where we used $\overline Z_0[\boldpsi={\mathbf 0},  \mathbf{h}]=1$ for all $\mathbf{h}$.

Further, the pairing $\langle \hat x_k(T)  x_l(T') \rangle_0 $ is zero for $k \neq l$ because the matrix of bare response functions is diagonal. We also note that the bare response function $R^0_{ii}(t,t')$ is only non-zero for $t>t'$ due to causality.  The vanishing of so-called `vacuum diagrams' and the significance of time-ordering is known as the Deker-Haake theorem \cite{deker1975fluctuation,dedominicis1978dynamics}. 

To find the average in the angular brackets in Eq.~(\ref{wickexample}) we must sum every possible product of pairwise averages of the dynamic variables. That is, we sum every possible pairing of the form
\begin{align}
A^1_{ij}(t,t') =& \frac{1}{ 2!} \left( \frac{\gamma \sigma^2}{N^2} \right)^2 \int dT dT' dT'' dT''' \sum_{(k,l,m),(k',l',m')}\Bigg[ \langle x_i(t) \hat x_m(T') \rangle_0 \langle x_l(T') \hat x_l(T)  \rangle_0 \nonumber \\
&\times \langle   x_k(T) \hat x_{m'}(T''')\rangle_0  \langle  x_{l'}(T''') \hat x_{l'}(T'')\rangle_0  \langle  x_{k'}(T'') \hat x_j(t')\rangle_0\Bigg]
\end{align}
The particular Wick pairing above can be represented by the following diagram 
\begin{figure}[H]
	\centering 
	\includegraphics[scale = 0.4]{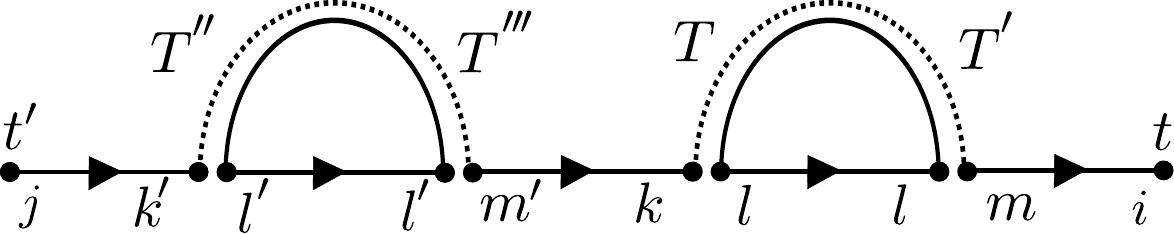}
\end{figure}
In this diagram, each variable (i.e. an $x$ or an $\hat x$) is represented by a small node. Nodes that are placed closely together share time coordinates. All the internal indices $k, l, m, k', l'$ and $m'$ are summed over. Indices associated with a pair of closely placed nodes are forbidden to take the same value in this summation (e.g. $l'$ and $m'$). Dashed undirected curves connect dynamic variables that are also forbidden from sharing a value (e.g. $k$ and $m$ in the above example). On the other hand, solid undirected curves connect dynamic variables that share an index. Finally, a directed horizontal edge points from a hatted variable to an unhatted variable, indicating that these are averaged as a pair (e.g. $x_{k}$ and $\hat x_{m'}$). Each directed edge therefore corresponds to a factor of the bare response function. 

\noindent The above diagram evaluates to (after carrying out sums over free indices)
\begin{align}
A^1_{ij}(t,t') = \frac{1}{2!} \left( \frac{\gamma \sigma^2}{N^2} \right)^2  \int dT dT' dT'' dT''' \sum_{l, k, l'} R^0_{ii}(t,T') R^0_{ll}(T',T)R^0_{kk}(T,T''') R^0_{l'l'}(T''',T'') R^0_{jj}(T'',t').
\end{align}
We note now that we could have considered a Wick pairing with the following exchange of indices and time coordinates: $k\leftrightarrow k', l\leftrightarrow l', m\leftrightarrow m', T\leftrightarrow T'', T'\leftrightarrow T'''$ [keeping the nodes $(i, t)$ and $(j,t')$ fixed]. In doing so, we would have obtained a diagram with the same structure and value. The number of ways of reordering the dashed and undashed coordinates is $2!$, cancelling the prefactor in Eq.(\ref{wickexample}). This is a general rule that holds for all diagrams. This means that we can drop the specific labellings in the inner nodes in the diagrams in conjunction with the factorials in the sum in Eq.~(\ref{interactionactionexpansion}). This simplifies our task of evaluating every possible Wick pairing to performing a sum over all the possible different architectures of diagram, ignoring diagrams that vanish trivially.

\subsection{Dressed response function and final expression for the outlier}\label{section:finalexpression}
Having introduced the diagrammatic formalism, we wish to find the dressed response functions for the case $\gamma \neq 0$ but with $r = c =  0$. We show in Section \ref{appendix:noeffect} that these additional type-2 correlations do not affect the response functions and consequently the outlier.

We begin with the expansion in Eq.~(\ref{interactionactionexpansion}). To lowest order in $S_{\mathrm{int}}$, the dressed and bare response functions are equal. Next, considering the term linear in $S_{\mathrm{int}}$, we obtain diagrams of the form
\begin{figure}[H]
	\centering 
	\includegraphics[scale = 0.4]{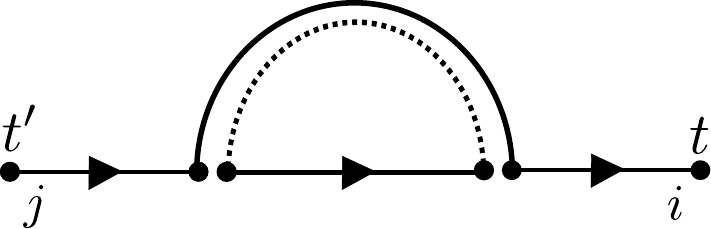}\\
	\vspace{4mm}
	\includegraphics[scale = 0.4]{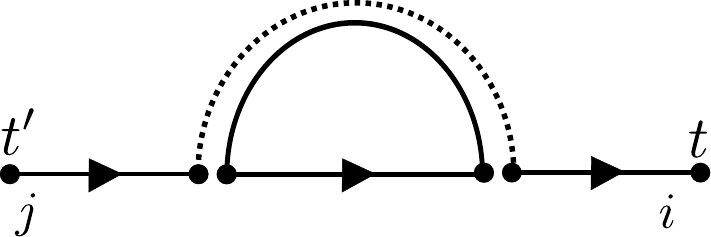}
	\captionsetup{labelformat=empty}
\end{figure}
These are the only first-order diagrams that do not vanish trivially as described in Section \ref{section:diagrams}. However, the first of these diagrams evaluates to zero since there are two nodes that are connected by a directed edge but simultaneously cannot share the same index value; because the bare resolvent is diagonal, this diagram must vanish. We omit further diagrams with similar architecture. The second diagram yields the following first-order approximation for the Laplace transform of the dressed response functions
\begin{align}
\hat R_{ij}(u) = \hat R^0_{ij}(u) +  \frac{\gamma \sigma^2}{N} \hat R^0_{ii}(u) \hat R^0(u) \hat R^0_{jj}(u) + \cdots ,
\end{align}
where $\hat R^0(u) = N^{-1} \sum_i \hat R^0_{ii}(u)$. Some example second-order diagrams in $S_{\mathrm{int}}$ are as follows 
\begin{figure}[H]
	\centering 
	\includegraphics[scale = 0.4]{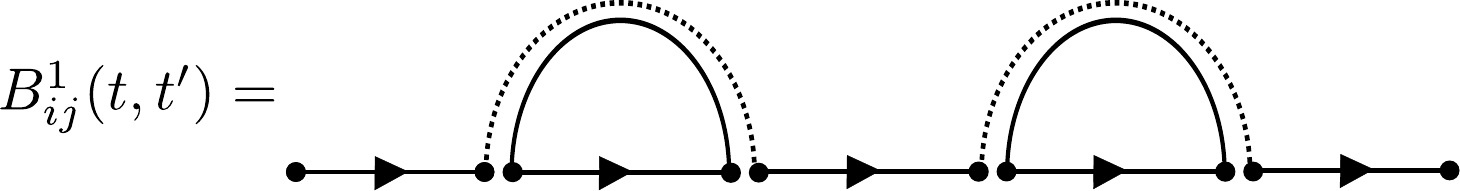}\\
	\vspace{4mm}
	\includegraphics[scale = 0.4]{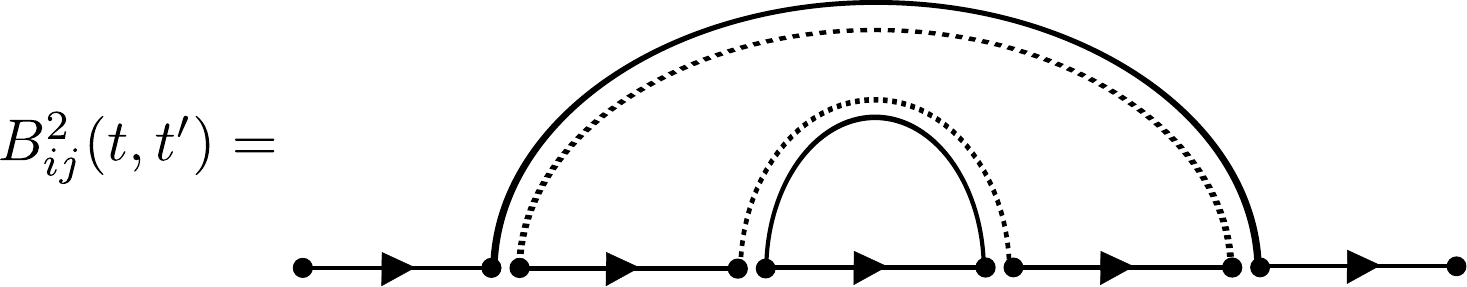}\\
	\vspace{4mm}
	\includegraphics[scale = 0.4]{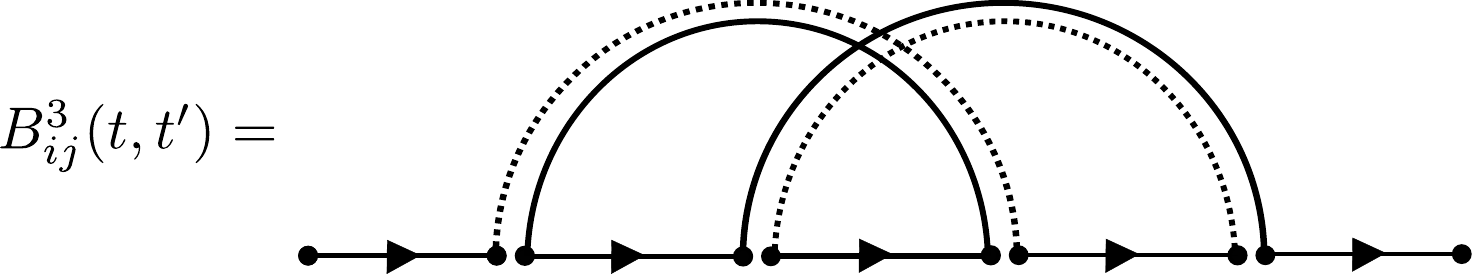}
	\captionsetup{labelformat=empty}
\end{figure}
\noindent Evaluating these diagrams we obtain
\begin{align}
\hat B^1_{ij}(u) &= \frac{1}{N}\left(\gamma \sigma^2\right)^2 \hat R^0_{ii}(u) \left[\hat R^0(u)\right]^3 \hat R^0_{jj}(u), \nonumber \\
\hat B^2_{ij}(u) &= \frac{1}{N}\left(\gamma \sigma^2\right)^2  \left[\hat R^0(u)\right]^3 \sum_k \hat R^0_{ik}(u) \hat R^0_{kj}(u), \nonumber \\
\hat B^3_{ij}(u) &= \frac{1}{N^3}\left(\gamma \sigma^2\right)^2 \hat R^0_{jj}(u) \hat R^0(u) \sum_{m,l,k} \hat R^0_{ik}(u)\hat R^0_{kl}(u)\hat R^0_{lm}(u).
\end{align}
When calculating the quantity $N^{-1}\sum_{ij}\hat R_{ij}(u)$ [which is crucial for the outlier eigenvalue -- see Eq.~(\ref{outliercalcgeneral})], we encounter the term $N^{-1}\sum_{ij}\hat B^1_{ij}(u)$ which is of the order $N^0$ and survives in the thermodynamic limit. However, the term $N^{-1}\sum_{ij}\hat B^2_{ij}(u)$ is of the order $N^{-1}$ and $N^{-1}\sum_{ij} \hat B^3_{ij}(u)$ is of the order $N^{-2}$, so that these two contributions vanish as $N\to\infty$. 

With these examples in mind, we can consider more generally which diagrams contribute in the thermodynamic limit. Because the bare resolvent is diagonal, the order of magnitude of the contribution of each diagram (in $N^{-1}$) to the sum $N^{-1} \sum_{ij} \hat R_{ij}(u)$ can be deduced from the number of separate groups of directed edges. By `group', we mean a set of directed edges that are connected by solid undirected arcs. This can be seen as follows:

In the first diagram above, there are five separate groups of undirected edges, giving rise to a factor of $N^5$ when all site indices are summed over. This is accompanied by a factor of $N^{-5}$, which arises from the square of the interaction term $S_{\mathrm{int}} \propto N^{-2}$ combined with the prefactor in the object $N^{-1}\sum_{ij}\hat B^1_{ij}(u)$. The term $N^{-1}\sum_{ij}\hat B^1_{ij}(u)$ is hence of order $N^0$. However, in the second diagram there are only four distinct groups of directed edges, leading to a factor $N^4$. Together with the same factor of $N^{-5}$ as before, this means that $N^{-1}\sum_{ij}\hat B^2_{ij}(u)$ is of order $N^{-1}$. Further, there are only 3 groups of undirected edges in the third diagram, meaning that $N^{-1}\sum_{ij}\hat B^3_{ij}(u)$ is of order $N^{-2}$. Hence,  $N^{-1}\sum_{ij}\hat B^2_{ij}(u)$ and $N^{-1}\sum_{ij}\hat B^3_{ij}(u)$ do not contribute in the thermodynamic limit. 

We can generalise this observation to higher orders. Accompanying the term $S_{\mathrm{int}}^r$ is a factor of $N^{-(2r+1)}$. At the $r^\mathrm{th}$ order, there are also $2r+1$ directed edges. The only way to construct a diagram at the $r^\mathrm{th}$ order with $2r+1$ disconnected directed edges is with an architecture like the diagram corresponding to $\hat B^1_{ij}(u)$. Similar deductions were used to justify neglecting so-called `non-planar' diagrams in the thermodynamic limit in other studies (known as 't Hooft's theorem in QCD \cite{kuczala2016eigenvalue, brezin1994correlation, HOOFT1974461}).

Thus, the series for the dressed response functions simplifies greatly and we obtain the diagrammatic series in Fig. 2 of the main text. This series evaluates to
\begin{align}
\hat R_{ij}& = \hat R^0_{ij} +  \frac{\gamma \sigma^2}{N} \hat R^0_{ii} \hat R^0 \hat R^0_{jj} + \frac{(\gamma \sigma^2)^2}{N} \hat R^0_{ii} (\hat R^0)^3 \hat R^0_{jj} + \frac{(\gamma \sigma^2)^3}{N} \hat R^0_{ii} (\hat R^0)^5 \hat R^0_{jj} + \cdots . \label{resolventserieswithnovel}
\end{align}
Evaluating the geometric series, we find 
\begin{align}
\frac{1}{N}\sum_{ij}\hat R_{ij}& = \hat R^0 \left[1 - \gamma \sigma^2 (\hat R^0)^2 \right]^{-1}. \label{sellresult}
\end{align}
We also note that the sum of only the diagonal elements is given by (in the thermodynamic limit)
\begin{align}
\frac{1}{N} \sum_i \hat R_{ii}& = \frac{1}{N} \sum_i \hat R^0_{ii} = \hat R^0, \label{autoresponsegamma}
\end{align}
with $\hat R^0$ as in Eq.~(\ref{response0}). That is, the inclusion of the interaction term preserves the average auto-response function, which is a requirement for our use of Wick's theorem (see Section \ref{appendix:quadraticaction}). 

We can now calculate the outlier in this case. Referring to Eq.~(\ref{outliercalcgeneral}) and substituting the expression in Eq.~(\ref{response0}) for $\hat R^0(u)$ into Eq.~(\ref{sellresult}), one obtains the result in Eq.~(14) of the main text, which we repeat here,
\begin{align}
\lambda_{\mathrm{outlier}}=
- 1 + \mu + \frac{\mu}{2}  \left(1 + \frac{\Gamma}{\gamma}\right)\left(\sqrt{1 + 4\gamma \sigma^2/\mu^2 }-1\right) . \label{outlierexpression}
\end{align}
where the expression is valid for $\vert \mu \vert > \sigma(1-\gamma)$ and there is no outlier for $\vert \mu \vert < \sigma(1-\gamma)$. It now remains to show two things: (1) that the contributions from correlations proportional to $r$ and $c$ can indeed be ignored, and (2) that the bare action can be replaced by the effective quadratic action in Eq.~(\ref{quadraticaction}). These points are addressed in subsections \ref{appendix:noeffect} and \ref {appendix:quadraticaction} respectively.

\subsection{In-row and in-column correlations have no effect on the outlier eigenvalue}\label{appendix:noeffect}
We have established in Section \ref{section:diagrams} that the correlations quantified by $\gamma$ affect the sum over all elements of the resolvent matrix $G_{ij}$, but have no effect on the diagonal elements of the resolvent matrix (or, equivalently, the auto-response functions). We now show that the remaining correlations quantified by $r$ and $c$ have no effect on the diagonal elements nor the sum $N^{-1}\sum_{ij}G_{ij}$. That is, they have no effect whatsoever on the eigenvalue spectrum.

Consider first the following contribution to the interaction term 
\begin{align}
S^{(1)}_{\mathrm{int}} &= -\frac{1}{2} \int dt dt' \bigg[ \frac{ r \sigma^2}{N^2} \sum_{(i,j,k)} \hat x_i(t) x_j(t) \hat x_i(t') x_k(t') \bigg] .
\end{align}
To first order, this gives rise only to diagrams of the form 
\begin{figure}[H]
	\centering 
	\includegraphics[scale = 0.3]{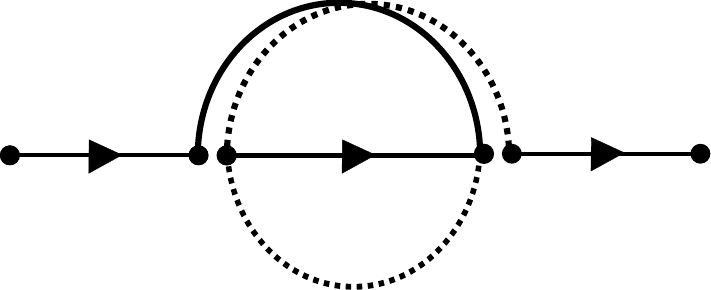}
\end{figure}
Since the bare response function matrix is diagonal, the first-order term proportional to $r$ vanishes. Considering now the second-order contributions, we obtain diagrams such as the following
\begin{figure}[H]
	\centering 
	\includegraphics[scale = 0.3]{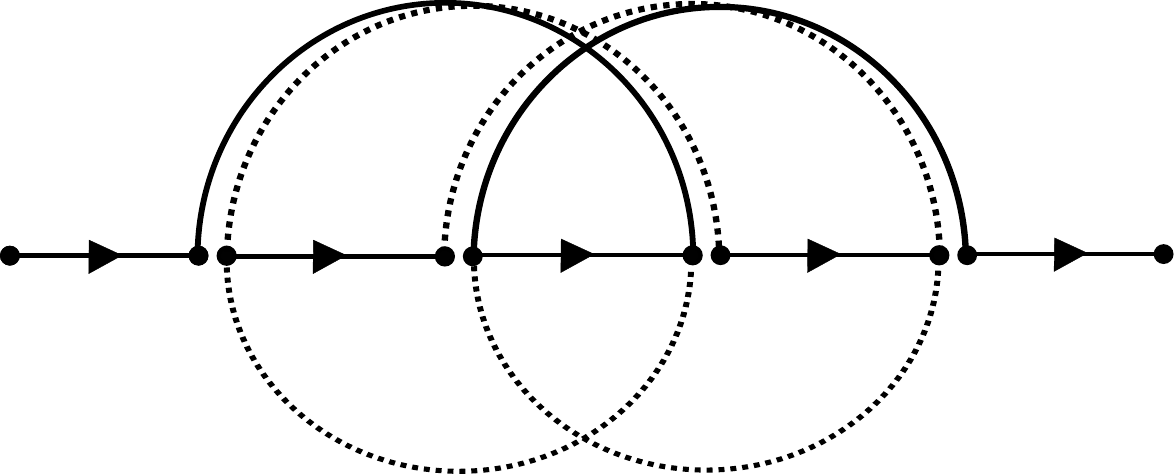}\\
	\vspace{4mm}
	\includegraphics[scale = 0.3]{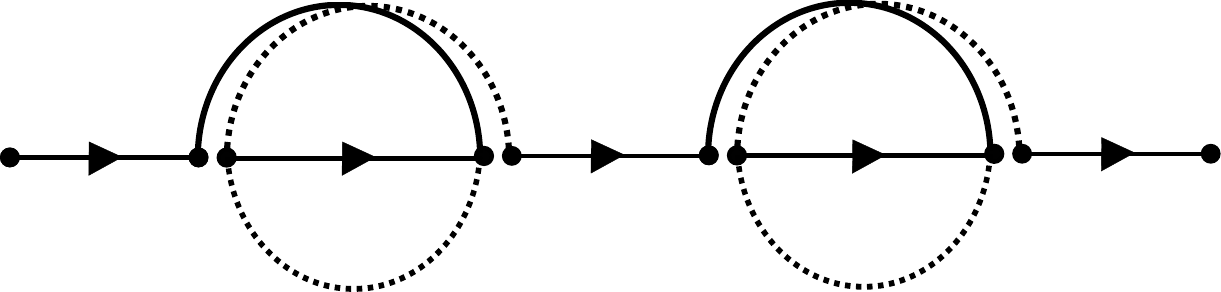}\\
	\vspace{4mm}
	\includegraphics[scale = 0.3]{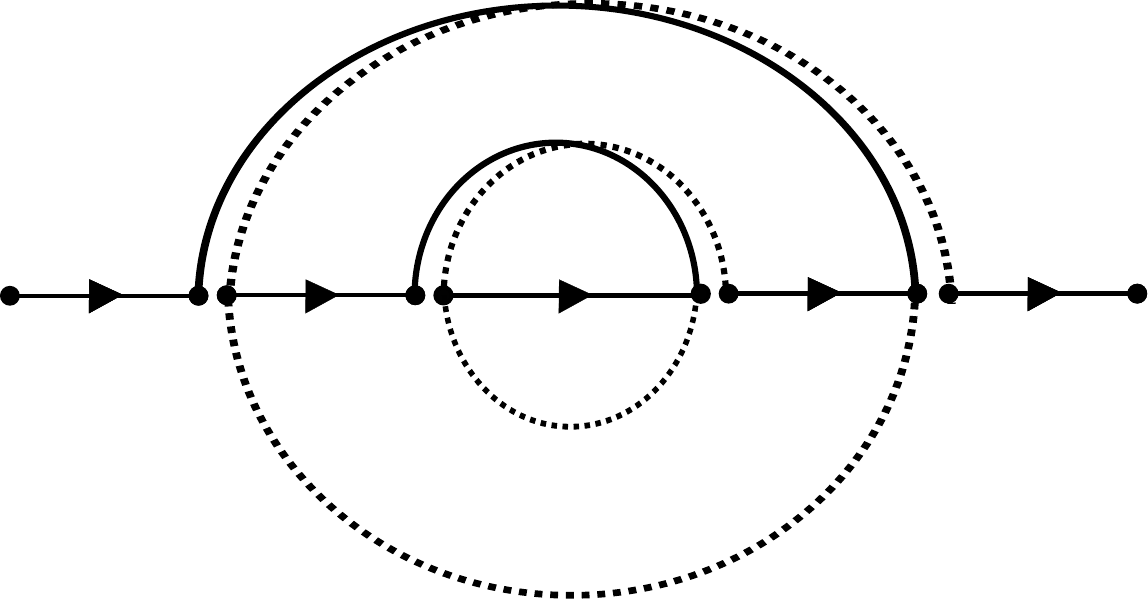}
	\captionsetup{labelformat=empty}
\end{figure}
The second and the last of these diagrams vanish once again. Following the reasoning in Section \ref{section:finalexpression}, we can deduce that the first diagram is of order $N^{-2}$ and so also vanishes in the thermodynamic limit. 

Similar deductions can be made for higher-order diagrams: We showed in Section \ref{section:finalexpression} that only diagrams that had all their directed edges disconnected from one another contributed in the thermodynamic limit. For this interaction term, no such diagrams are possible. Therefore, the interaction term proportional to $r$ does not have any effect on the eigenvalue spectrum in the thermodynamic limit. 

A very similar argument can be made about the following interaction term as well
\begin{align}
S_{\mathrm{int}}^{(2)} &= -\frac{1}{2} \int dt dt' \bigg[  \frac{ c \sigma^2}{N^2} \sum_{(i,j,k)} \hat x_i(t) x_j(t) \hat x_k(t') x_j(t') \bigg] .
\end{align}
Hence, in-row and and in-column correlations have no effect on the response functions. Therefore these correlations do not affect the outlier eigenvalue.

\subsection{Justification for the use of Wick's theorem -- saddle-point integration to obtain a quadratic action} \label{appendix:quadraticaction}
We show in this section how the generating functional can be manipulated so as to make the bare action $S_0$ in Eq.~(\ref{bareandinteraction}) quadratic. This means that we can proceed with the calculation of the dressed response function, using the expansion in Eq.~(\ref{interactionactionexpansion}), as if Wick's theorem still holds despite the bare action being originally quartic in the variables $x_i$ and $\hat x_i$. 

We begin by introducing
\begin{align}
C(t,t') &= \frac{1}{N} \sum_{i} x_i(t) x_i(t'), \nonumber \\
L(t,t') &= \frac{1}{N} \sum_{i} \hat x_i(t) \hat x_i(t'), \nonumber \\
K(t,t') &= \frac{1}{N} \sum_{i} x_i(t) \hat x_i(t').
\end{align}
Following Refs. \cite{dedominicis1978dynamics} we impose these definitions through Dirac-delta functions in the form of complex exponentials in the disorder-averaged generating functional in Eq.~(\ref{genfunctlinearprocess}), recalling  the action in Eq.~(\ref{bareandinteraction}). We then have
\begin{align}
\overline{Z[\boldpsi,  \mathbf{h}]} &= \int DC D\hat C DL D\hat L DK D\hat K \int D[\bx,\widehat\bx] \nonumber \\
& \exp\left(i\sum_i\int dt~ x_i(t)\psi_i(t)\right) \exp\left(S_\mathrm{int}[\bx,\widehat\bx]\right)\exp\Bigg(i\sum_i\int dt \Bigg[\widehat x_i(t)\Bigg(\dot x_i(t)+x_i(t)-h_i(t)\Bigg)\Bigg]\Bigg)\nonumber\\
&\times \exp\Bigg(-i\sum_i\int dt dt' \Bigg[ \hat C(t,t') x_i(t) x_i(t') + \hat L(t,t') \hat x_i(t) \hat x_i(t') + \hat K(t,t') x_i(t) \hat x_i(t')\Bigg]\Bigg) \nonumber \\
&\times\exp\Bigg(-\frac{N}{2} \int dt dt' \bigg[\sigma^2 C(t,t')L(t,t') + \Gamma \sigma^2  K(t,t')K(t',t) \bigg] \Bigg) \nonumber \\
&\times\exp\Bigg( i N \int dt dt' \bigg[ \hat C(t,t')C(t,t') +  \hat L(t,t')L(t,t') +  \hat K(t,t')K(t,t') \bigg] \Bigg).
\end{align}
One then carries out a saddle-point integration in the limit $N\to\infty$, and finds \begin{align}
C(t,t') &= \frac{1}{N}\sum_i \langle x_i(t) x_i(t')\rangle_\Omega, \nonumber \\
L(t,t') &= \frac{1}{N}\sum_i \langle \hat x_i(t) \hat x_i(t')\rangle_\Omega = 0, \nonumber \\
K(t,t') &= \frac{1}{N}\sum_i \langle x_i(t) \hat x_i(t')\rangle_\Omega, \nonumber \\
i \hat C(t,t') &= \frac{\sigma^2}{2} L(t,t') = 0 ,\nonumber \\
i \hat L(t,t') &= \frac{\sigma^2}{2} C(t,t')  ,\nonumber \\
i \hat K(t,t') &= \Gamma\sigma^2  K(t',t),
\end{align}
where 
\begin{align}
\langle \cdot \rangle_\Omega &= \frac{\int D[\bx,\widehat\bx] \left[ \cdot \right] e^\Omega}{\int D[\bx,\widehat\bx]  e^\Omega}. \nonumber \\
e^\Omega &= \exp\left( S_0^\mathrm{eff}[\bx,\widehat\bx]+ S_\mathrm{int}[\bx,\widehat\bx] \right), \label{omegaav}
\end{align}
where $R^0(t,t') = -i K(t,t')$ and $S_0^\mathrm{eff}[\bx,\widehat\bx]$ is given in Eq.~(\ref{quadraticaction}). After the saddle-point integration, the generating functional can be written as 
\begin{align}
\overline{Z[\boldpsi,  \mathbf{h}]} &= \mathcal{N} \int D[\bx,\widehat\bx] \exp\left( S_0^\mathrm{eff}[\bx,\widehat\bx]+ S_\mathrm{int}[\bx,\widehat\bx] + i\sum_i\int dt~ x_i(t)\psi_i(t)  \right), \label{genfunctaftersaddle}
\end{align}
where $\mathcal{N}$ is a normalisation constant. Eq.~(\ref{genfunctaftersaddle}) (with $S_{\mathrm{int}} = 0$) then describes the effective stochastic process defined self-consistently in Eqs.~(\ref{effectiveprocess}) and (\ref{sc}). Crucially, the action for this effective process $S_0^\mathrm{eff}[\bx,\widehat\bx]$ is quadratic in the variables $\{x_i(t),\hat x_i(t)\}$. This justifies the use of Wick's theorem for averages against the bare action. The bare response function can be found directly from the effective process by first functionally differentiating then taking the Laplace transform. One obtains the expression in Eq.~(\ref{bareresponseeq}). 

Because of self-consistent nature of the effective process in Eq.~(\ref{effectiveprocess}) and because the interaction term enters into the average in Eq.~(\ref{omegaav}), Eq.~(\ref{bareresponseeq}) cannot be used unless the dressed auto-response functions are the same as the bare auto-response functions. That is to say, we can assume an effective quadratic bare action in the thermodynamic limit (and therefore that Wick's theorem holds) only if the auto-response function, which corresponds to the average diagonal element of the resolvent matrix, is unaffected by the inclusion of the interaction term. This is something we must check \textit{a posteriori}. This does indeed turn out be true in the present case [see Eq.~(\ref{autoresponsegamma})]. 

\subsection{Piecewise nature of the bare response function}\label{appendix:constraints}
From the self-consistent process in Eqs.~(\ref{effectiveprocess}) and (\ref{sc}) associated with the effective bare action, we deduced that the bare response function must satisfy the quadratic equation in Eq.~(\ref{bareresponseeq}). This equation has two solutions, but the response function is single-valued. We wish to decide which of these solutions is the correct one. 

Let us consider the Laplace transform of the process in Eq.~(\ref{effectiveprocess})
\begin{align}
\left[ 1+u - \Gamma \sigma^2 \hat R^0(u)\right] \hat x_i(u) &= x_i(0)+ \hat \eta_i(u) .
\end{align}
Multiplying both sides by their complex conjugates, taking an average over realisations of the noise and using Eq.~(\ref{bareresponseeq}), one finds
\begin{align}
\langle \vert \hat x_i(u) \vert^2\rangle_0 = \frac{\vert x_i(0) \vert^2}{\vert \hat R^0(u)\vert^{-2} - \sigma^2 }.
\end{align}
In order for this equation to be consistent, we must have that the quantity on the right-hand side is greater than zero. This means that
\begin{align}
\vert \hat R^0(u)\vert^{2} <  \frac{1}{\sigma^2 }.
\end{align}
Using this condition, the correct solution to Eq.~(\ref{bareresponseeq}) for each value of $u$ can be deduced, yielding Eq.~(\ref{response0}).

\section{The bulk spectrum}\label{section:bulk}

One cannot use the same series expansion techniques that were used to calculate the outlier eigenvalue to evaluate the resolvent in the bulk region of the eigenvalue spectrum. This is because the resolvent is no longer an analytic function in the bulk region. Fortunately, one only requires the diagonal elements of the resolvent matrix to evaluate the bulk eigenvalue density, so we can use more traditional methods in this case.

Here, we employ a method that was originally developed by Sommers et al \cite{sommers} to show that the type-2 correlations do not effect the bulk eigenvalue spectrum in the limit $N\to \infty$ so that the familiar elliptic law is recovered.

\subsection{The eigenvalue potential}
For completeness, we summarise some helpful results from previous works Refs. \cite{sommers, baron2020dispersal}. We define the eigenvalue potential of the ensemble of (real) random matrices $\underline{\underline{a}}$ as
\begin{align}
\Phi(\omega) = -\frac{1}{N} \overline{\ln \det\left[ (\id \omega^\star - \underline{\underline{a}}^T )( \id \omega - \underline{\underline{a}} )\right]} . \label{potential}
\end{align}
This potential is related to the trace of the resolvent matrix $G(\omega, \omega^\star) = \frac{1}{N} \sum_i G_{ii}(\omega, \omega^\star)$ via
\begin{align}
G(x,y) = \frac{\partial \Phi(\omega, \omega^\star)}{\partial \omega}, \label{resolventtracefromphi}
\end{align}
where $\omega = x + i y$. The eigenvalue density $\rho$ is obtained via
\begin{align}
\frac{\partial \mathrm{Re} [G]}{\partial x} - \frac{\partial \mathrm{Im} [G]}{\partial y} = \mathrm{Re} \left[ \frac{\partial G(\omega, \omega^\star)}{\partial \omega^\star}\right] = 2\pi \rho .
\end{align}
Therefore, when the trace of the resolvent matrix is analytic, the eigenvalue density must be zero. 

\subsection{Replica method and Hubbard-Stratonovich transformation}
We note that in order to evaluate the potential in Eq.~(\ref{potential}), we need to perform the ensemble average of a logarithmic quantity. To do this we use the replica method (see e.g. \cite{mezard1987}). This involves writing the logarithm as $\ln x = \lim_{n\to 0} \frac{1}{n} (x^n -1)$. 

Using a Gaussian integral, we can write an $n$-fold replicated version of the determinant in Eq.~(\ref{potential}) 
\begin{align}\
&\det\left[ (\id\omega^\star - \underline{\underline{a}}^T )(\id\omega - \underline{\underline{a}} )\right]^{-n} = \int \prod_{i, \alpha} \left( \frac{d^2z^\alpha_i}{\pi}\right) \nonumber \\
&\times \exp\left[ -\epsilon \sum_i \vert z^\alpha_i \vert^2 - \sum_{i,j,k, \alpha}z^{\alpha \star}_i (\omega^\star\delta_{ik} - (a^T)_{ik})(\omega\delta_{kj} - a_{kj})z^\alpha_j\right] \nonumber \\
&=\int \prod_{i\alpha} \left( \frac{d^2z^\alpha_i d^2y^\alpha_i}{2 \pi^2}\right) \exp\left[ - \sum_{i\alpha} y^{\alpha\star}_i y^\alpha_i + \epsilon z^{\alpha\star}_i z^\alpha_i \right]\nonumber \\
\times &\exp\left[ i \sum_{ij\alpha} z^{\alpha\star}_i \left( a_{ji} - \omega^\star\delta_{ij} \right) y^\alpha_j\right] \exp\left[ i \sum_{ij \alpha} y^{\alpha\star}_i \left( a_{ij} - \omega \delta_{ij}\right) z^\alpha_j\right], \label{HStransformed}
\end{align}
where $\epsilon$ is a positive infinitesimal quantity introduced so as to avoid singularities which occur when $\omega$ is equal to one of the eigenvalues of $\underline{\underline{a}}$. In the second line we have performed a Hubbard--Stratonovich transformation \cite{hubbard} in order to linearise the terms in the exponential involving $a_{ij}$.
 
Following for example Edwards and Jones \cite{edwardsjones}, one can show that after taking the disorder average of this quantity, the entire exponent can be written as the sum of $n$ replicated terms in the limit $N\to \infty$. That is, the replicas decouple and we can write
\begin{align}
\overline{\det\left[ (\id\omega^\star - \underline{\underline{a}}^T )(\id\omega - \underline{\underline{a}} )\right]^{-n}} = \left[\overline{\det\left[ (\id\omega^\star - \underline{\underline{a}}^T )(\id\omega - \underline{\underline{a}} )\right]}\right]^{-n},
\end{align}
and thus the following holds (as was noted in other works \cite{haake, sommers})
\be\label{eq:help}
\overline{\ln \det\left[ (\id\omega^\star -\underline{\underline{a}}^T )( \id\omega - \underline{\underline{a}} )\right]}= \ln \overline{\det\left[ (\id\omega^\star - \underline{\underline{a}}^T )( \id\omega - \underline{\underline{a}} )\right]}.
\ee
We arrive at the following expression for the eigenvalue potential after taking the disorder average [in a similar fashion to Eq.~(\ref{disorderaverage})]
\begin{align}
\exp\left[ -N \Phi(\omega) \right] &= \int \prod_{i} \left( \frac{d^2z_i d^2y_i}{2 \pi^2}\right) \exp\left[ - \sum_{i} y^{\star}_i y_i + \epsilon z^{\star}_i z_i \right]\nonumber \\
\times & \exp\left[ -i \sum_{i } z^{\star}_i y_i  (\omega^\star + 1) + z_i y^{\star}_i ( \omega + 1) + \frac{i \mu}{N} \sum_{ij}  z^{\star}_i y_j + z_j y^{\star}_i   \right]\nonumber \\
\times & \exp\left[ - \frac{\sigma^2}{2N}\sum_{i j} \left(z_i^{\star} y_j + z_i y_j^{\star}\right)\left(z_i^{\star} y_j + z_i y_j^{\star}\right) + \Gamma \left(  z_i^{\star} y_j + z_i y_j^{\star}\right)\left( z_j^{\star} y_i + z_j y_i^{\star}\right) \right] \nonumber \\
\times & \exp\left[ - \frac{\gamma\sigma^2}{N^2}\sum_{(i,j,k) } \left(z_i^{\star} y_j + z_i y_j^{\star}\right)\left(z_k^{\star} y_i + z_k y_i^{\star}\right)  \right] \nonumber \\
\times & \exp\left[ - \frac{\sigma^2}{2N^2}\sum_{(i,j,k) } r \left(z_i^{\star} y_j + z_i y_j^{\star}\right)\left(z_i^{\star} y_k + z_i y_k^{\star}\right) + c \left(z_i^{\star} y_j + z_i y_j^{\star}\right)\left(z_k^{\star} y_i + z_k y_i^{\star}\right)  \right] . \label{averaged2}
\end{align}

\subsection{Independence of the bulk eigenvalue density from type-2 correlations and non-zero mean}\label{section:independence}

We wish to evaluate the integral in Eq.~(\ref{averaged2}) in the limit $N \to \infty$ and thus deduce the bulk eigenvalue spectrum. To this end, we now introduce the order parameters
\begin{align}
u &= \frac{1}{N} \sum_{i } z^{\star}_i z_i, \,\,\,\,\, v = \frac{1}{N} \sum_{i } y^{\star}_i y_i, \nonumber \\
w &= \frac{1}{N} \sum_i z^{\star}_i y_i, \,\,\,\,\, w^\star = \frac{1}{N} \sum_i y^{\star}_i z_i, \label{orderparameters}
\end{align}
which we impose in the integral Eq.~(\ref{averaged2}) using Dirac delta functions in their complex exponential representation. We can thus rewrite Eq.~(\ref{averaged2}) as 
\begin{align}
\exp\left[ -N \Phi(\omega) \right] &= \int \mathcal{D}\left[ \cdots \right]\exp\left[N(\Psi + \Theta + \Omega)\right] ,\label{saddlesetup}
\end{align}
where where $\mathcal{D}\left[ \cdots \right]$ denotes integration over all of the order parameters and their conjugate (`hatted') variables, and where 
\begin{align}
\Psi &= i  \hat u u  + i\hat v  v  + i \hat w  w^\star  +   i \hat w^\star  w  , \nonumber \\
\Theta &= -\epsilon  u - v - \sigma^2 u v    -i w (\omega^\star + 1)  - i w^\star (\omega + 1)  - \frac{1}{2} \sigma^2 \Gamma [w^2+ ( w^\star )^2] , \nonumber \\
\Omega &= N^{-1}\ln  \left[\int   \prod_{i} \left( \frac{d^2z_i d^2y_i}{2 \pi^2}\right)  \exp\left\{ -i \sum_i  \left(\hat u z_i^{\star} z_i + \hat v y_i^{\star} y_i  + \hat w y_i^{\star} z_i  + \hat w^\star z_i^{\star} y_i  \right) \right\} \exp\left[ \chi\left(\{z_i, z^\star_i, y_i, y^\star_i\}\right) \right]\right]. \label{psithetaomega}
\end{align}
The quantity $\chi\left(\{z_i, z^\star_i, y_i, y^\star_i\}\right)$ contains all the terms in the exponent of Eq.~(\ref{averaged2}) that are proportional to any of the following quantities:  $N^{-1} \sum_i (y_i)^2$, $N^{-1} \sum_i (z_i)^2$, $N^{-1} \sum_i z_i y_i $, $N^{-1} \sum_i  y_i $, $N^{-1} \sum_i z_i $ and their complex conjugates.

We wish to show now that, in the saddle-point approximation, the order parameters in Eq.~(\ref{orderparameters}) are unaffected by the terms in $\chi\left(\{z_i, z^\star_i, y_i, y^\star_i\}\right)$. In particular, we wish to show that $iw^\star$, which in the saddle-point approximation corresponds to the trace of the resolvent [see Eq.~(\ref{resolventtracefromphi}) and Refs.~\cite{haake, baron2020dispersal}, is unaffected by setting $\chi\left(\{z_i, z^\star_i, y_i, y^\star_i\}\right) = 0$.

Extremising the quantity $\Psi + \Theta + \Omega$ with respect to the conjugate order parameters $\hat u$, $\hat v$, $\hat w$ and $\hat w^\star$, we obtain saddle-point conditions of the form
\begin{align}
w^\star &= N^{-1} \sum_{i} \left\langle y^{\star}_i z_i \right\rangle_\Omega, \nonumber \\
\left\langle \cdots  \right\rangle_\Omega &= \frac{1}{\Omega} \int   \prod_{i} \left( \frac{d^2z_i d^2y_i}{2 \pi^2}\right) \left[ \cdots\right]  \exp\left\{ -i \sum_i  \left(\hat u z_i^{\star} z_i + \hat v y_i^{\star} y_i  + \hat w y_i^{\star} z_i  + \hat w^\star z_i^{\star} y_i  \right) \right\} \nonumber \\
&\;\;\;\;\;\;\;\;\;\;\;\;\;\;\;\;\;\;\;\;\;\times\exp\left[ \chi\left(\{z_i, z^\star_i, y_i, y^\star_i\}\right) \right],
\end{align}
with similar expressions for the other order parameters. Taking inspiration from our earlier calculation of the response functions in Section \ref{section:outlier},  we expand the exponent containing $\chi$ as a series and evaluate the resulting series for $\left\langle y^{\star}_i z_i \right\rangle_\Omega$ as a set of Gaussian integrals. One finds that all the non-zero-order terms in $\chi$ in the series vanish in the thermodynamic limit. This is true for all the other order parameters in Eq.~(\ref{orderparameters}) also. Similar observations were made in Refs. \cite{sommers, haake, edwardsjones}. One can thus set $\chi = 0$ in the expression for $\Omega$. This means that the terms proportional $\mu$, $\gamma$, $r$ and $c$ do not contribute in the thermodynamic limit to the bulk spectrum. 

Once $\chi$ is discarded, one finds
\begin{align}
\Omega = - \ln(\hat w \hat w^\star - \hat u \hat v ).
\end{align}
The expression in Eq.~(\ref{psithetaomega}) is then precisely the form that was derived in \cite{anandgalla}, from which the familiar elliptic law can be obtained. Therefore, neither the non-zero mean nor the type-2 correlations affect the bulk spectrum in the thermodynamic limit and we obtain Girko's elliptic law [given in Eq.~(7) of the main text].

\clearpage

\section{Generating random matrices with general correlations} \label{section:procedure}

\subsection{Procedure}
In order to test the predictions for the bulk spectrum and outlier eigenvalue, we developed an efficient method for generating random matrices with the statistics given in Eq.~(2) of the main text.

We decompose $z_{ij}$ (for $i\neq j$) as, 
\begin{equation}
z_{ij}=\beta_i+\kappa_j+\omega_{ij},
\label{eq:sampling-fneq0-decomposition}
\end{equation}
where $\beta_i$, $\kappa_j$, and $\omega_{ij}$ are zero-mean random variables that are correlated in such a way as to reproduce the desired correlations in $z_{ij}$. More specifically, we require
\begin{align}
\begin{bmatrix}
\langle \beta_i^2\rangle & \langle \beta_i\kappa_i\rangle \\
\langle \beta_i\kappa_i\rangle & \langle \kappa_i^2\rangle \\
\end{bmatrix} &=
\frac{\sigma^2}{N^2}\begin{bmatrix}
r & \gamma\\
\gamma & c \\
\end{bmatrix},
\label{eq:sampling-fneq0-cov-beta-kappa}\\
\begin{bmatrix}
\langle \omega_{ij}^2\rangle & \langle \omega_{ij}\omega_{ji}\rangle \\
\langle \omega_{ij}\omega_{ji}\rangle & \langle \omega_{ij}^2\rangle \\
\end{bmatrix} &= \frac{\sigma^2}{N^2}\begin{bmatrix}
1-\frac{r+c}{N} & \Gamma-\frac{2\gamma}{N} \\
\Gamma-\frac{2\gamma}{N} & 1-\frac{r+c}{N} \\
\end{bmatrix},
\label{eq:sampling-fneq0-cov-omega}
\end{align}
where all other correlations are nil. Using these covariances, we sample $N$ correlated pairs $(\beta_i,\kappa_i)$, and a further  $\frac{N(N-1)}{2}$ correlated pairs $(\omega_{ij},\omega_{ji})$. We can then combine the $\{\beta_i, \kappa_j, \omega_{ij}\}$ using Eq.~(\ref{eq:sampling-fneq0-decomposition}) to produce a single realisation of the matrix $\underline{\underline{z}}$.

In order for this method to work, the covariance matrices in equation Eqs.~(\ref{eq:sampling-fneq0-cov-beta-kappa}) and (\ref{eq:sampling-fneq0-cov-omega}) must be positive semi-definite. This implies the following conditions,
\begin{align}
r &\geq 0,\nonumber \\
c &\geq 0,\nonumber \\
 \vert \gamma \vert &\leq \sqrt{rc},\nonumber\\
r+c &\leq N,\nonumber\\
\left\vert\Gamma - \frac{2 \gamma}{N} \right\vert &\leq 1-\frac{r + c}{N}.
\end{align}
These conditions are more stringent than the constraints given in Section \ref{section:generalconstraints}. This is due to the decomposition of $z_{ij}$ chosen in Eq.~(\ref{eq:sampling-fneq0-decomposition}), which, at the cost restricting the values of the parameters we can test, makes the production of matrices with generalised correlations very efficient. 

\subsection{Non-Gaussian distributions used in Fig. 3}
In order to verify both the accuracy of our prediction for the eigenvalue spectrum and its universal nature, we draw the elements of $z_{ij}$ from two different non-Gaussian distributions in Fig.~3 of the main paper.

Consider two correlated random variables $x$ and $y$, both with zero mean, which have the following statistics
\begin{align}
\begin{bmatrix}
\langle x^2\rangle & \langle xy\rangle \\
\langle xy\rangle & \langle y^2\rangle \\
\end{bmatrix} &= \begin{bmatrix}
v_x& c_{xy} \\
c_{xy} & v_y \\
\end{bmatrix}.
\label{covmatxy}
\end{align}
The variables $x$ and $y$ will have the above statistics if they are drawn from the following distribution, 
\begin{align}
P(x, y ) &= \frac{1}{2\sqrt{3 v_x}}\theta\left( \sqrt{3 v_x} - x\right)\theta\left(x + \sqrt{3v_x}\right)\nonumber \\
&\times \left[\frac{(1+c_{xy}/\sqrt{v_xv_y})}{2 } \delta\left( \sqrt{\frac{v_y }{v_x}} x - y\right)  + \frac{(1-c_{xy}/\sqrt{v_xv_y})}{2 } \delta\left(\sqrt{\frac{v_y }{v_x}} x + y\right)\right], \label{uniform_distribution}
\end{align}
but the marginal distribution of either variable will be a uniform distribution.

All the correlated pairs of variables $(\beta_i,\kappa_i)$ and $(\omega_{ij},\omega_{ji})$ [see Eq.~(\ref{eq:sampling-fneq0-decomposition})] were drawn from such a distribution [with statistics given by Eqs.~(\ref{eq:sampling-fneq0-cov-beta-kappa}) and (\ref{eq:sampling-fneq0-cov-omega}))] to produce the points in Fig.~3(b) in the main text. 

Similarly, the following distribution also gives the same statistics as Eq.~(\ref{covmatxy}), but the marginal distribution is a Bernoulli distribution 
\begin{align}
P(x, y) &= \frac{1 +c_{xy}/\sqrt{v_xv_y} }{4} \left[\delta(x - \sqrt{v_x})\delta( y - \sqrt{v_y}) + \delta(x + \sqrt{v_x})\delta(y + \sqrt{v_y}) \right] \nonumber \\
&+ \frac{(1-c_{xy}/\sqrt{v_xv_y})}{4 } \left[\delta(x - \sqrt{v_x})\delta(y + \sqrt{v_y}) + \delta(x + \sqrt{v_x})\delta(y - \sqrt{v_y}) \right]. \label{bernoulli_distribution}
\end{align}
This distribution was used to produce the points in Fig.~3(a) in the main text.